\documentclass[useAMS,usenatbib,usegraphicx]{mn2e}

\title[Kinematics of OB-associations] {Kinematics of OB-associations  in Gaia epoch}
\author[A.M. Mel'nik and A.K. Dambis]{A.M.Mel'nik\thanks{E-mail: anna@sai.msu.ru}   and A.K.
Dambis\\ Sternberg Astronomical Institute, Lomonosov Moscow State
University, Universitetskii pr. 13, Moscow, 119991 Russia}

\begin{document}

\date{Accepted 2017 August 23. Received 2017 December 23; in original form 2017 April 11}

%\pagerange{\pageref{firstpage}--\pageref{lastpage}} \pubyear{2009}

\maketitle

\label{firstpage}

\begin{abstract}
We use stellar proper motions from the  TGAS catalog  to study the
kinematics of OB-associations. The TGAS proper motions of
OB-associations generally agree well with the Hipparcos proper
motions. The parameters of the Galactic rotation curve obtained
with TGAS and Hipparcos proper motions agree within the errors.
The average one-dimensional  velocity dispersion inside 18
OB-associations with more than 10 TGAS stars is $\sigma_v=3.9$ km
s$^{-1}$, which is considerably smaller, by a factor of 0.4, than
the velocity dispersions derived from Hipparcos data.  The
effective contribution from orbital motions of binary OB-stars
into the velocity dispersion $\sigma_v$ inside OB-associations is
$\sigma_b=1.2$ km s$^{-1}$. The median virial and stellar masses
of OB-associations are equal to 7.1 10$^5$ and 9.0 10$^3$
M$_\odot$, respectively. Thus OB-associations must be unbound
objects provided they do not include  a lot of dense gas. The
median star-formation efficiency is $\epsilon=2.1$ percent. Nearly
one third of stars of OB-associations must lie outside their tidal
radius.  We found that the Per OB1 and Car OB1 associations are
expanding with the expansion started in a small region of 11--27
pc 7--10 Myr ago. The average expansion velocity is 6.3 km
s$^{-1}$.

\end{abstract}


\section{Introduction}

Galactic astronomy is on the threshold of a new era of high
precision proper motions with the recent first {\it Gaia} data
release \citep[{\it Gaia} DR1,][]{gaia2016b} from the European
Space Agency (ESA). {\it Gaia} DR1 contains the {\it Tycho-Gaia}
Astrometric Solution \citep[TGAS,][]{michalik2015,lindegren2016}
which provides positions, parallaxes and proper motions for about
2 million stars using the 24 yr time difference between Hipparcos
\citep{esa1997, hog2000} and {\it Gaia} \citep{gaia2016a}
observations.

The term 'OB-association' was first introduced by
\citet{ambartsumian1949}. The sizes of OB-associations differ from
10 pc (Cyg OB2) to 500 pc (Cep OB1), although the sky-plane sizes
of 90 percent of them do not exceed 200 pc
\citep{blahahumphreys1989}. They often contain young clusters in
their centres \citep{garmany1992} and sometimes have several
centres of concentration \citep{melnik1995}. The catalog compiled
by \citet{blahahumphreys1989} includes 91 OB-associations located
within 3.5 kpc from the Sun. This catalog may also include several
young open clusters because it  is difficult to make a clear
distinction between OB-associations and young open clusters.
Though, OB-associations have, on average, larger sizes: 80 percent
of them are larger than  $d>20$ pc. Moreover, OB-associations are
less centrally concentrated than young open clusters.

There are several partitions of high-luminosity stars (OB-stars
and red supergiants) into OB-associations
\citep{humphreys1984,blahahumphreys1989, garmany1992, melnik1995}.
\citet{garmany1992}  divided young stars into  OB-associations in
the sector of Galactic longitudes 55--150$^\circ$.
\citet{melnik1995} used cluster analysis method to identify the
densest and most compact parts of OB-associations, but these
groups include only  few stars with known kinematical data. Both
lists of OB-associations are based on  photometric data obtained
by \citet{blahahumphreys1989}. Here we consider the partition of
\citet{blahahumphreys1989} as the most universal.

The catalog by \citet{blahahumphreys1989} includes the massive end
of the stellar population of OB-associations. In some other
studies  \citep[for example,][]{dezeeuw1999}, the authors are
concerned with the entire stellar population and also consider
smaller scales, such as subgroups within larger OB-associations.

OB-associations are supposed to form in  giant molecular clouds
\citep[][and references therein]{elmegreen1983, zinnecker2007}.
The diameters and masses of giant molecular clouds lie in the
range 10--80 pc and  10$^5$ -- 2 10$^6$ M$_\odot$
\citep{sanders1985}, respectively. The catalog of Galactic giant
molecular clouds is expected to be essentially complete at $M>3$
10$^5$ M$_\odot$ \citep{solomon1987,solomon1989}. There is a lot
of evidence that  giant molecular clouds are  close to virial
equilibrium \citep{larson1981,krumholz2006}.

The efficiency of star formation, $\epsilon$, determined as the
ratio of the stellar mass of OB-association to the gaseous mass of
its parent giant molecular clouds   usually lies  in the range of
0.1--10 percent \citep{myers1986, evans2009, garcia2014}. The low
efficiency of star formation can be explained by two different
ways: through destruction of  molecular clouds by the radiation of
high luminosity stars \citep[for example,][]{elmegreen1983,
franco1994, colin2013} and through supersonic turbulence and
magnetic fields preventing global collapse of the cloud
\citep{maclow2004, mckee2007}.

\citet{blaauw1964}  found the expansion of OB-associations by
analyzing ground-based proper motions of their member stars.
\citet{brown1997} simulated the expansion of OB-associations to
check methods for deriving their kinematic age -- the time in the
past when the OB-association had minimal size. \citet{madsen2002}
find some evidence for expansion in Sco OB2 by comparing
spectroscopic radial velocities with values derived from the
hypothesis that members of the OB-association share the same
velocity vector. Although the estimated expansion velocities  of
OB-associations decrease with growing accuracy of proper motions,
the expansion of these systems is expected in many scenarios of
star formation.

The mass loss in a gas cloud due to  thermal pressure of HII
regions \citep{mckee1989,kim2016} can make the newly formed
stellar group unbound. If mass is ejected from the system within a
time comparable to the crossing time, then the system becomes
unbound after 50 percent mass loss \citep{hills1980}. However,
more accurate treatment of  relaxation processes shows that the
system can form an expanding OB-association with a bound cluster
in the centre
\citep{tutukov1978,kroupa2001,boily2003a,boily2003b,vine2003,
baumgardt2007}.

We study the kinematics of OB-associations  using the stellar
proper motions from the  TGAS catalog \citep{michalik2015}. The
high precision of TGAS proper motions makes it possible to
estimate the virial masses of OB-associations corresponding to the
masses of their parent molecular clouds, and examine the expansion
of OB-associations at present time.

In section 2 we investigate the motions of OB-associations as single
entities: we derive the parameters of the Galactic rotation curve,
study the residual velocities and the motion along the Z-axis.  In
section 3 we study internal properties of OB-associations:
we determine their velocity dispersions,  virial and stellar masses,
estimate the star formation efficiency and tidal radii.
We consider the expansion of OB-associations in Section 4 and
formulate the main conclusions in Section 5.

\section{Motions of OB-associations in the Galaxy}

\subsection{Kinematic data of stars in OB-associations}

We calculate the median  proper motions of OB-associations using
TGAS proper motions of individual stars \citep{michalik2015}. On
the whole, OB-associations identified by
\citet{blahahumphreys1989} include 500 stars with known TGAS
proper motions. For comparison, Hipparcos catalog contains 774
star members of OB-associations. The lack of TGAS proper motions
mostly concerns nearby OB-associations, which contain bright stars
($m_v<7^m$). Determining astrometric parameters of bright stars
requires additional calibrations and for this reason TGAS lacks
many bright stars from nearby OB-associations. For example,
OB-association Ori OB1 includes 59 stars with known Hipparcos
proper motions and only three stars with TGAS data. We therefore
consider only proper motions of stars in OB-associations and do
not use  their parallaxes. It is nearby stars with the most
precise parallaxes that are most important for the study of the
distance scale.

The mass measurements of proper motions of blue stars became
available due to the inclusion of a list of OB-stars into the
Hipparcos input catalog  \citep{dezeeuw1999} providing very
accurate first-epoch positions and hence very small TGAS
proper-motion errors for OB-association members. The average error
of TGAS proper motions for OB-association member stars is
$\varepsilon_{\mu l}=\varepsilon_{\mu b}=0.059$ mas yr$^{-1}$,
which is nearly 15 times smaller than the average error of
Hipparcos proper motions, 0.916 mas yr$^{-1}$, for OB-association
stars.

Table~\ref{tgas_data} presents the average Galactic coordinates of
OB-associations, $l$ and $b$, the average heliocentric distances
to the associations, $r$, the median proper motions in l- and
b-directions, $\mu_l$ and $\mu_b$, the dispersion of proper
motions  $\sigma_{\mu l}$ and $\sigma_{\mu b}$ and the number of
stars in OB-associations with known TGAS proper motions, $n_\mu$.
To complete the kinematic data we also give the median line-of
sight velocities, $V_r$, the  dispersion of the line-of-sight
velocities $\sigma_{vr}$ and the number of stars with known
line-of-sight velocities, $n_{vr}$, taken from the catalog by
\citet{barbierbrossat2000}. Table 1 also gives the total number of
stars in an OB-association with known photometric measurements,
$N_t$, used by \citet{blahahumphreys1989} to determine distances
for OB-associations, $r_{BH}$.

Note that the errors of the listed  proper motions and
line-of-sight velocities of OB-associations, $\varepsilon_{\mu
l}$, $\varepsilon_{\mu b}$ and $\varepsilon_{vr}$, are calculated
in the following way:
\begin{equation}
\varepsilon_{\mu l}=\frac{\sigma_{\mu l}}{\sqrt{n_\mu}},
\label{err_ml}
\end{equation}

\begin{equation}
\varepsilon_{\mu b}=\frac{\sigma_{\mu b}}{\sqrt{n_\mu}},
\label{err_mb}
\end{equation}

\begin{equation}
\varepsilon_{vr}=\frac{\sigma_{vr}}{\sqrt{n_{vr}}}.
 \label{err_vr}
\end{equation}

\noindent The method of deriving  the robust estimates of the
dispersion of $\mu_l$, $\mu_b$ and $V_r$ inside the OB-association
is also discussed in section 3.1.

Table 2, which is available in the online version of the paper,
gives the spectral, photometric and kinematic data of stars in
OB-associations. It presents the name of a star, the name of the
OB-association to which it is assigned by
\citet{blahahumphreys1989}, spectral type  of the star, code of
its luminosity class $c_L$: 2 -- Ia,  4 -- Iab, 6 -- Ib, 8 -- II,
10 -- III, 12 -- IV, 14 -- V, where the corresponding odd numbers
(1, 3, ..., 13) reflect the uncertainty in its determination.
Table 2 also shows Galactic coordinates $l$ and $b$ of a star, and
the heliocentric distance $r$ to OB-association, which it is
assigned to. We also present the line-of-sight velocity of a star
$V_r$ and its error $\varepsilon_{vr}$ taken from the catalog
\citet{barbierbrossat2000}. Note that we use only the individual
stellar velocities $V_r$ determined with the error
$\varepsilon_{vr}\le 10$ km s$^{-1}$ to derive the median values
of $V_r$ and velocity dispersions $\sigma_{vr}$ in OB-association.
For the  Hipparcos stars we present their Hipparcos number
$n_\textrm{Hip}$, TGAS proper motions, $\mu_l$ and $\mu_b$, if
avalable, and their errors, $\varepsilon_{\mu l}$ and
$\varepsilon_{\mu b}$. If a Hipparcos star is absent in the TGAS
catalog, then we give its Hipparcos proper motions and their
errors; flag F indicates the source of proper motions: 'G' means
TGAS and 'H' -- Hipparcos. Table 2 also represents  color indexes
$B-V$ and $U-B$, apparent and absolute magnitudes, $m_V$ and
$M_V$, and the $V$-band extinction, $A_V$, that are adopted from
the catalog by \citet{blahahumphreys1989}.

We reduced the heliocentric distances to OB-associations derived
by \citet{blahahumphreys1989}, $r_{BH}$,  to the short distance
scale, $r=0.8\cdot r_{BH}$ \citep{sitnik1996,
dambis2001,melnikdambis2009}.  We also correct the absolute
stellar magnitudes obtained by \citet{blahahumphreys1989},
$M_{V(BH)}$, to bring them onto  the short distance scale
$M_V$=$M_{V(BH)}$+$\Delta m$, where $\Delta m=-5\log 0.8=0.485^m$.

%--------------------------- Table 1 ----------------------------------------
\begin{table*}
\caption{Line-of-sight velocities and TGAS proper motions of
OB-associations}
 \begin{tabular}{lrrrrrrrrrrrr}
  \hline
 Name & $l\quad$ & $b\;\;$ & $r\;\;$ & $N_t$ & $V_r\;\;$ & $\sigma_{vr}\;\;$ & $n_{vr}$ &
 $\mu_l\quad$ & $\sigma_{\mu l} \quad\;$ & $\mu_b\quad$ & $\sigma_{\mu b} \quad\;$ & $n_{\mu}$\\
  & deg. & deg. & kpc &  & km s$^{-1}$ & km s$^{-1}$ &  & mas yr$^{-1}$ & mas yr$^{-1}$ & mas yr$^{-1}$ & mas yr$^{-1}$ \\
 \hline
SGR OB5   &   0.02 &  -1.19 &  2.42 &  31 &  -15.0 &  19.0 &   2 &   -0.35 &        &   -2.39 &        &   1  \\
SGR OB1   &   7.55 &  -0.77 &  1.26 &  65 &  -10.0 &  12.1 &  37 &   -1.10 &   0.22 &   -1.30 &   0.78 &  13  \\
SGR OB7   &  10.73 &  -1.57 &  1.39 &   4 &   -6.1 &  17.2 &   3 &         &        &         &        &   0  \\
SGR OB4   &  12.11 &  -0.96 &  1.92 &  15 &    3.5 &  10.7 &   9 &   -0.24 &   0.79 &   -1.41 &   0.34 &   2  \\
SGR OB6   &  14.19 &   1.28 &  1.60 &   5 &   -7.3 &   0.1 &   4 &         &        &         &        &   0  \\
SER OB1   &  16.72 &   0.07 &  1.53 &  43 &   -5.0 &  20.0 &  17 &   -1.10 &   0.63 &   -0.87 &   0.34 &   6  \\
SCT OB3   &  17.30 &  -0.73 &  1.33 &  10 &    3.3 &  17.0 &   8 &   -2.51 &   0.61 &   -1.07 &   0.64 &   3  \\
SER OB2   &  18.21 &   1.63 &  1.60 &  18 &   -4.0 &  14.5 &   7 &   -0.98 &   0.11 &   -0.16 &   0.08 &   2  \\
SCT OB2   &  23.18 &  -0.54 &  0.80 &  13 &  -11.0 &  20.0 &   6 &   -0.51 &   0.34 &   -0.74 &   0.04 &   4  \\
TR 35     &  28.03 &  -0.46 &  2.01 &   9 &   31.0 &       &   1 &   -3.05 &        &   -0.93 &        &   1  \\
COLL 359  &  29.79 &  12.63 &  0.16 &   1 &        &       &   0 &         &        &         &        &   0  \\
VUL OB1   &  60.35 &   0.03 &  1.60 &  28 &    3.1 &  14.7 &   9 &   -4.98 &   1.10 &   -0.86 &   0.52 &   8  \\
VUL OB4   &  60.63 &  -1.22 &  0.80 &   9 &   -2.9 &   7.5 &   3 &   -5.10 &   0.21 &   -1.74 &   0.65 &   2  \\
CYG OB3   &  72.77 &   2.03 &  1.83 &  42 &   -9.5 &   9.5 &  30 &   -7.15 &   0.30 &   -0.89 &   0.18 &  16  \\
CYG OB1   &  75.84 &   1.12 &  1.46 &  71 &  -13.5 &  10.5 &  34 &   -6.35 &   0.54 &   -0.58 &   0.36 &  12  \\
CYG OB9   &  77.81 &   1.80 &  0.96 &  32 &  -19.5 &   8.7 &  10 &   -6.44 &   1.13 &   -1.95 &   1.22 &   5  \\
CYG OB8   &  77.91 &   3.36 &  1.83 &  21 &  -21.0 &  11.0 &   9 &   -6.27 &   0.43 &    0.53 &   1.60 &  10  \\
CYG OB2   &  80.27 &   0.88 &  1.46 &  15 &        &       &   0 &   -5.24 &        &   -0.80 &        &   1  \\
CYG OB4   &  82.69 &  -7.48 &  0.80 &   2 &   -4.9 &   1.1 &   2 &         &        &         &        &   0  \\
CYG OB7   &  88.99 &   0.03 &  0.63 &  29 &   -9.4 &   9.3 &  21 &   -2.43 &   3.77 &   -1.01 &   0.68 &  16  \\
NGC 6991  &  87.58 &   1.42 &  1.39 &   1 &  -15.0 &       &   1 &         &        &         &        &   0  \\
LAC OB1   &  96.71 & -17.70 &  0.48 &   2 &  -13.6 &   4.2 &   2 &         &        &         &        &   0  \\
CEP OB2   & 102.04 &   4.68 &  0.73 &  57 &  -17.0 &   6.7 &  36 &   -3.84 &   0.91 &   -0.77 &   1.23 &  34  \\
CEP OB1   & 104.20 &  -0.94 &  2.78 &  58 &  -58.2 &   7.4 &  17 &   -4.46 &   0.69 &   -0.67 &   0.36 &  20  \\
NGC 7235  & 102.78 &   0.77 &  3.18 &   1 &        &       &   0 &         &        &         &        &   0  \\
CEP OB5   & 108.50 &  -2.69 &  1.67 &   6 &  -48.7 &  29.2 &   2 &   -4.17 &        &   -0.47 &        &   1  \\
CAS OB2   & 111.99 &  -0.00 &  2.10 &  41 &  -50.1 &  11.0 &   7 &   -4.02 &   0.07 &   -0.59 &   0.31 &   5  \\
CEP OB3   & 110.42 &   2.56 &  0.70 &  26 &  -22.9 &   3.9 &  18 &   -2.13 &   0.51 &   -1.71 &   0.40 &  13  \\
CAS OB5   & 116.10 &  -0.50 &  2.01 &  52 &  -45.8 &   7.2 &  16 &   -3.72 &   0.86 &   -0.86 &   0.41 &   8  \\
CEP OB4   & 118.21 &   5.25 &  0.66 &   7 &  -24.0 &       &   1 &   -1.84 &   0.05 &   -1.29 &   0.04 &   4  \\
CAS OB4   & 120.05 &  -0.30 &  2.30 &  27 &  -37.0 &   8.6 &   7 &   -2.50 &   0.17 &   -0.92 &   0.13 &   6  \\
CAS OB14  & 120.37 &   0.74 &  0.88 &   8 &  -15.0 &   7.0 &   4 &   -0.87 &   0.08 &   -1.21 &   0.99 &   2  \\
CAS OB7   & 122.98 &   1.22 &  2.01 &  39 &  -50.0 &   1.0 &   4 &   -2.47 &   0.23 &   -0.47 &   0.24 &   7  \\
CAS OB1   & 124.72 &  -1.73 &  2.01 &  11 &  -42.0 &   2.5 &   5 &   -1.51 &   0.20 &   -1.27 &   0.27 &   3  \\
NGC 457   & 126.64 &  -4.43 &  2.01 &   4 &  -34.8 &   2.3 &   4 &   -1.55 &   0.06 &   -0.93 &   0.02 &   2  \\
CAS OB8   & 129.16 &  -1.06 &  2.30 &  43 &  -34.6 &   9.9 &  14 &   -1.00 &   0.10 &   -0.61 &   0.15 &   8  \\
PER OB1   & 134.70 &  -3.16 &  1.83 & 165 &  -43.2 &   7.0 &  80 &   -0.19 &   0.58 &   -1.23 &   0.36 &  58  \\
CAS OB6   & 135.02 &   0.75 &  1.75 &  46 &  -42.6 &   8.1 &  12 &   -0.38 &   0.29 &   -0.74 &   0.47 &  11  \\
CAM OB1   & 141.08 &   0.89 &  0.80 &  50 &  -11.0 &   9.4 &  30 &    0.13 &   1.25 &   -1.27 &   0.92 &  26  \\
CAM OB3   & 146.99 &   2.85 &  2.65 &   8 &  -27.6 &  19.3 &   3 &   -0.70 &        &    0.35 &        &   1  \\
PER OB3   & 146.64 &  -5.86 &  0.14 &   1 &   -2.4 &       &   1 &         &        &         &        &   0  \\
PER OB2   & 160.24 & -16.55 &  0.32 &   7 &   21.2 &   4.5 &   7 &    5.60 &   0.15 &   -1.16 &   3.89 &   3  \\
AUR OB1   & 173.83 &   0.14 &  1.06 &  36 &   -1.9 &  14.0 &  26 &    2.25 &   0.51 &   -1.57 &   0.30 &  12  \\
ORI OB1   & 206.94 & -17.71 &  0.40 &  68 &   25.4 &   7.9 &  62 &   -0.03 &   2.28 &   -0.30 &   0.59 &   3  \\
AUR OB2   & 173.33 &  -0.17 &  2.42 &  20 &   -2.6 &   4.9 &   4 &    0.85 &   0.84 &   -0.79 &   0.30 &   2  \\
NGC 1893  & 173.60 &  -1.70 &  2.90 &  10 &        &       &   0 &         &        &         &        &   0  \\
NGC 2129  & 186.45 &  -0.11 &  1.46 &   3 &   16.8 &   7.5 &   3 &    1.90 &        &   -0.91 &        &   1  \\
GEM OB1   & 188.98 &   2.22 &  1.21 &  40 &   16.0 &   5.0 &  18 &    1.53 &   0.39 &   -0.56 &   0.31 &   6  \\
MON OB1   & 202.10 &   1.08 &  0.58 &   7 &   23.4 &  13.0 &   7 &    0.27 &   0.32 &   -1.62 &   0.18 &   4  \\
MON OB2   & 207.46 &  -1.65 &  1.21 &  32 &   22.3 &  12.5 &  26 &   -1.10 &   0.22 &   -1.31 &   0.47 &  10  \\
MON OB3   & 217.65 &  -0.44 &  2.42 &   4 &   27.0 &       &   1 &         &        &         &        &   0  \\
CMA OB1   & 224.58 &  -1.56 &  1.06 &  17 &   34.3 &  16.2 &   8 &   -2.93 &   0.78 &   -1.93 &   1.15 &   7  \\
NGC 2414  & 231.09 &   1.01 &  3.18 &  15 &   67.2 &       &   1 &   -1.94 &        &   -0.50 &        &   1  \\
COLL 121  & 238.45 &  -8.41 &  0.55 &  13 &   29.6 &   7.0 &  10 &   -4.73 &   1.77 &   -1.53 &   0.37 &   3  \\
NGC 2362  & 237.87 &  -5.92 &  1.21 &   9 &   30.0 &  14.0 &   6 &         &        &         &        &   0  \\
NGC 2367  & 235.65 &  -3.84 &  2.20 &   5 &   37.0 &   9.0 &   4 &   -3.11 &        &   -0.66 &        &   1  \\
NGC 2439  & 245.27 &  -4.08 &  3.50 &  23 &   62.7 &       &   1 &   -4.27 &   0.43 &   -0.54 &   0.18 &  10  \\
PUP OB1   & 243.53 &   0.16 &  2.01 &  22 &   77.0 &       &   1 &   -3.66 &   0.11 &   -0.94 &   1.64 &   3  \\
PUP OB2   & 244.61 &   0.58 &  3.18 &  13 &        &       &   0 &         &        &         &        &   0  \\
COLL 140  & 244.47 &  -7.33 &  0.29 &   6 &   10.3 &   6.6 &   5 &   -7.56 &   1.23 &   -4.45 &   1.55 &   2  \\
 \hline
\label{tgas_data}
\end{tabular}
\end{table*}
%------------------------------------------------------------------------------
\addtocounter{table}{-1}
%--------------------------- Table 1 ----------------------------------------

\begin{table*}
\caption{Line-of-sight velocities and TGAS proper motions of
OB-associations (end)}
 \begin{tabular}{lrrrrrrrrrrrr}
  \hline
 Name & $l\quad$ & $b\;\;$ & $r\;\;$ & $N_t$ & $V_r\;\;$ & $\sigma_{vr}\;\;$ & $n_{vr}$ &
 $\mu_l\quad$ & $\sigma_{\mu l} \quad \;$ & $\mu_b\quad$ & $\sigma_{\mu b}\quad \;$ & $n_{\mu}$\\
  & deg. & deg. & kpc &  & km s$^{-1}$ & km s$^{-1}$ &  & mas yr$^{-1}$ & mas yr$^{-1}$ & mas yr$^{-1}$ & mas yr$^{-1}$ \\
 \hline
PUP OB3   & 253.89 &  -0.25 &  1.46 &   3 &        &       &   0 &         &        &         &        &   0  \\
VELA OB2  & 262.08 &  -8.52 &  0.39 &  13 &   24.0 &   9.7 &  13 &   -9.73 &   0.32 &   -4.88 &   2.97 &   2  \\
VELA OB1  & 264.84 &  -1.41 &  1.46 &  46 &   23.0 &   4.3 &  18 &   -6.78 &   1.07 &   -1.69 &   0.25 &   7  \\
CAR OB1   & 286.45 &  -0.46 &  2.01 & 126 &   -5.0 &   8.2 &  39 &   -8.08 &   0.74 &   -0.81 &   0.24 &  15  \\
TR 16     & 287.25 &  -0.25 &  2.10 &  18 &   -1.0 &   3.3 &   5 &   -7.20 &   0.55 &   -0.94 &   0.38 &   2  \\
TR 14     & 287.37 &  -0.47 &  2.78 &   5 &  -10.0 &       &   1 &         &        &         &        &   0  \\
TR 15     & 287.65 &  -0.42 &  3.04 &   2 &        &       &   0 &         &        &         &        &   0  \\
COLL 228  & 287.57 &  -0.98 &  2.01 &  15 &  -13.0 &   9.0 &   9 &   -6.92 &        &   -1.64 &        &   1  \\
CAR OB2   & 290.39 &   0.12 &  1.79 &  59 &   -8.2 &   8.5 &  22 &   -6.62 &   0.23 &   -1.09 &   0.27 &  10  \\
NGC 3576  & 291.33 &  -0.61 &  2.53 &   5 &  -17.0 &   9.0 &   2 &         &        &         &        &   0  \\
CRU OB1   & 294.87 &  -1.06 &  2.01 &  76 &   -5.3 &   8.9 &  33 &   -6.28 &   0.32 &   -0.88 &   0.21 &  17  \\
NGC 3766  & 294.11 &  -0.02 &  1.53 &  12 &  -15.6 &   0.7 &   2 &   -6.76 &   0.01 &   -1.10 &   0.09 &   2  \\
CEN OB1   & 304.14 &   1.44 &  1.92 & 103 &  -19.0 &  14.5 &  32 &   -4.89 &   0.57 &   -1.01 &   0.21 &  28  \\
HOGG 16   & 307.51 &   1.39 &  1.46 &   5 &  -35.0 &   8.5 &   3 &   -3.70 &   0.01 &   -1.09 &   0.24 &   2  \\
R 80      & 309.37 &  -0.41 &  2.90 &   2 &  -38.2 &       &   1 &         &        &         &        &   0  \\
NGC 5606  & 314.87 &   0.99 &  1.53 &   5 &  -37.8 &   1.8 &   3 &   -5.60 &   0.04 &   -0.85 &   0.01 &   2  \\
CIR OB1   & 315.47 &  -2.76 &  2.01 &   4 &        &       &   0 &         &        &         &        &   0  \\
PIS 20    & 320.39 &  -1.49 &  3.18 &   6 &  -49.0 &       &   1 &   -4.98 &        &   -0.20 &        &   1  \\
NOR OB1   & 328.05 &  -0.92 &  2.78 &   8 &  -35.6 &   6.5 &   6 &   -4.11 &        &   -0.69 &        &   1  \\
NGC 6067  & 329.71 &  -2.18 &  1.67 &   9 &  -40.0 &   2.6 &   8 &   -3.18 &        &   -0.57 &        &   1  \\
R 103     & 332.34 &  -0.75 &  3.18 &  34 &  -48.0 &  26.0 &  11 &   -5.04 &   1.38 &   -1.52 &   0.73 &   3  \\
R 105     & 333.08 &   1.90 &  1.26 &   4 &  -31.0 &   7.0 &   4 &   -3.04 &        &    0.07 &        &   1  \\
ARA OB1B  & 337.95 &  -0.85 &  2.78 &  21 &  -34.7 &  10.3 &   9 &   -2.63 &   0.64 &   -1.20 &   0.43 &   6  \\
ARA OB1A  & 337.69 &  -0.92 &  1.10 &  53 &  -36.3 &  20.6 &   8 &   -2.05 &   0.33 &   -2.59 &   0.12 &   5  \\
NGC 6204  & 338.31 &  -1.15 &  2.20 &  14 &  -51.0 &   5.8 &   5 &         &        &         &        &   0  \\
SCO OB1   & 343.71 &   1.36 &  1.53 &  76 &  -28.8 &  17.5 &  28 &   -2.26 &   0.41 &   -0.56 &   0.32 &   8  \\
SCO OB2   & 351.31 &  19.02 &  0.13 &  10 &   -4.1 &   2.3 &  10 &         &        &         &        &   0  \\
HD 156154 & 351.30 &   1.41 &  2.10 &   4 &   -4.0 &   8.5 &   3 &   -1.05 &   0.10 &   -0.54 &   0.09 &   2  \\
SCO OB4   & 352.64 &   3.23 &  0.96 &  11 &    3.0 &   6.3 &   7 &   -1.34 &   0.42 &   -2.70 &   0.29 &   3  \\
TR 27     & 355.06 &  -0.73 &  0.88 &  11 &  -15.8 &       &   1 &         &        &         &        &   0  \\
M 6       & 356.75 &  -0.87 &  0.37 &   1 &   -6.4 &       &   1 &         &        &         &        &   0  \\
 \hline
\label{mu_data}
\end{tabular}
\end{table*}
%------------------------------------------------------------------------------
%--------------------------- Table 2 ------------------------------------------
\begin{table*}
\caption{Spectral, photometric and kinematic data for stars  in
OB-associations (available online only)}
\end{table*}
%------------------------------------------------------------------------------

\subsection{Comparison  TGAS and Hipparcos proper motions of OB-associations}

\begin{figure*}
\resizebox{15 cm}{!}{\includegraphics{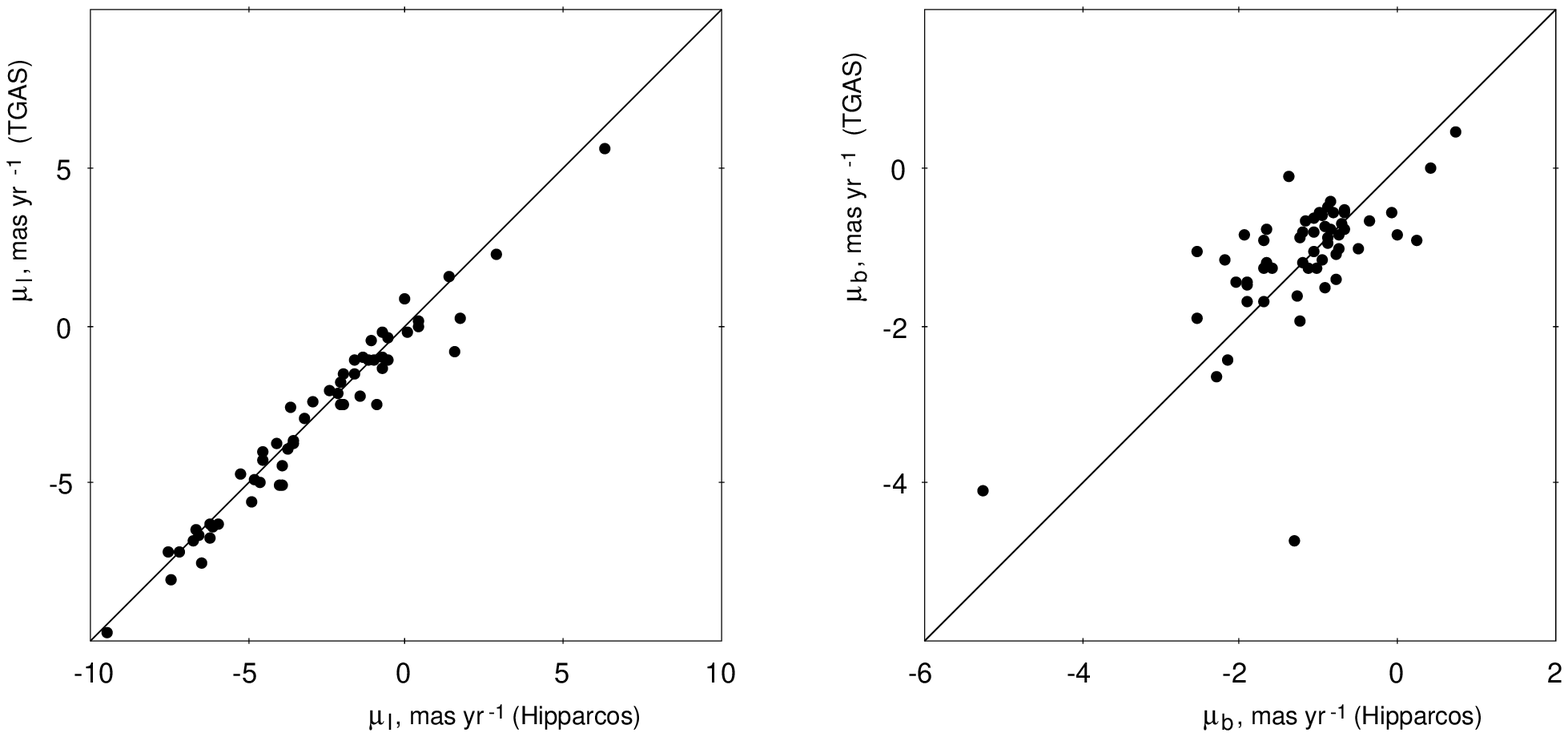}}
\caption{Comparison of   proper motions $\mu_l$ and $\mu_b$ of
OB-associations derived from data of Hipparcos and TGAS catalogs.
Only 56 OB-associations containing at least two stars with known
proper motions are considered. Note that the left and right images
have different scales. The rms deviation of TGAS and Hipparcos
proper motions is the same for l- and b-direction and equal to
$\sigma_{\mu}=0.67$ mas yr$^{-1}$.} \label{mu_com}
\end{figure*}

We compare the median proper motions $\mu_l$ and $\mu_b$ of
OB-associations derived from data of Hipparcos and TGAS catalogs.
Only 56 OB-associations containing at least two stars with known
proper motions are considered. Figure~\ref{mu_com} shows the
correlation between the Hipparcos and TGAS proper motions. The
vast interval spanned by $\mu_l$ values, [-10,+6] mas yr$^{-1}$,
is due to differential Galactic rotation. The distribution of
proper-motion components $\mu_b$ is more compact, reflecting small
motions perpendicular to the Galactic plane. Two associations with
large negative $\mu_b$ (Coll 140 and Vela OB2) are located close
to the Sun, $r=0.3$ and 0.4 kpc, so their large negative proper
motions translate into the velocities of only -7 and -9 km
s$^{-1}$. The $\mu_b$ proper-motion components are mostly negative
reflecting the positive velocity of the solar motion perpendicular
to the Galactic plane, $w_0$, which is equal to about +7 km
s$^{-1}$. The rms difference of TGAS and Hipparcos proper motions
is the same for l- and b-direction and equal to
$\sigma_{\mu}=0.67$ mas yr$^{-1}$. This value is comparable to the
average error of proper motions of OB-stars in the Hipparcos
catalog, 0.916 mas yr$^{-1}$. Thus the mean difference of
velocities calculated at the mean heliocentric distance of
OB-associations ($r=1.5$ kpc) is $\Delta v=4.7$ km s$^{-1}$.

%-----------------------    Figure 1   ----------------------------------------
%------------------------------------------------------------------------------

\subsection{Galactic rotation curve}

We determine the parameters of the Galactic rotation curve assuming
to  a first approximation that OB-associations move in circular
orbits in accordance with Galactic  differential rotation. The
method of the determination of the parameters of the rotation curve is
described in detail in \citet{melnikdambis2009}.

Briefly, we expand the angular rotation velocity $\Omega(R)$ into
a power series in $R-R_0$ to the second order and determine values
of three parameters of the rotation curve: the angular velocity at
the solar distance $\Omega_0$ as well as the first and second
derivatives of $\Omega(R)$ taken at the solar distance,
$\Omega'_0$ and $\Omega''_0$, respectively. We also determine the
solar velocity components $u_0$ and $v_0$  with respect to the
centroid of OB-associations in the direction toward the Galactic
centre and in the sense of Galactic rotation, correspondingly.

We use two samples of OB-associations (denoted as samples 1 and
2). Sample 1 includes objects whose median line-of-sight
velocities $V_r$ or proper motions $\mu_l$ were derived from the
data for at least two member stars, whereas sample 2 is based on
the velocities and proper motions obtained from the data for at
least five member  stars. Sample 1 provides 70 and 56 equations
for $V_r$ and $\mu_l$, respectively. Sample 2 gives 50 equations
for $V_r$ and 32 equations for $\mu_l$. We solve the equations for
the line-of-sight velocities and proper motions jointly and use
weight factors  to allow for observational errors and "cosmic"
velocity dispersion \citep[see also][]{dambis1995,
melnik1999,melnik2001}. We use standard least square method
\citep{press1987} to solve the systems of 126 (sample 1) and 82
(sample 2)  equations, which are linear in the parameters
$\Omega_0$, $\Omega'_0$, $\Omega''_0$, $u_0$, and $v_0$.

We adopt a solar Galactocentric distance of $R_0=7.5$ kpc
\citep{rastorguev1994, dambis1995, glushkova1998,
nikiforov2004,feast2008,groenewegen2008,reid2009b,dambis2013,
francis2014,boehle2016,branham2017}. Note that the particular
choice of $R_0$ in the range 7--9 kpc has practically no effect on
the analysis of the space distribution and kinematics of stars
located within 3 kpc from the Sun.

Table~\ref{tab_rot_curve} lists the  parameters of the Galactic
rotation curve, $\Omega_0$, $\Omega'_0$ and $\Omega''_0$, and the
solar motion toward the apex, $u_0$ and $v_0$, determined for
samples 1 and 2. It also lists the inferred Oort constant
$A=-0.5R_0\Omega'_0$, the standard deviation of the velocities
from the rotation curve $\sigma_0$, and the number of conditional
equations $N_{eq}$.

We can see that the parameters of the Galactic rotation curve and
solar motion to the apex derived for samples 1 and 2 agree within
the errors. They also agree well with  the parameters obtained for
Hipparcos data \citep[see Table 2 in][] {melnikdambis2009}. The
greatest difference between the parameters derived from TGAS and
Hipparcos proper motions  concerns the case of the second
derivative of the angular velocity $\Omega(R)$ taken at the solar
distance, $\Omega''_0$: $\Omega''_0=1.11\pm0.22$ and $1.35\pm0.20$
km s$^{-1}$ kpc$^{-3}$, respectively. Note that both TGAS and
Hipparcos data yield  large values for the angular velocity at the
solar  distance $\Omega_0$ equal to $31.08\pm0.86$ and
$30.55\pm0.87$ km s$^{-1}$ kpc$^{-1}$, respectively. The
$\Omega_0=31.08$ km s$^{-1}$ kpc$^{-1}$ estimate translates into a
solar azimuthal velocity of $\Theta=233$ km s$^{-1}$. A large
$\Omega_0$ value of 30--31 km s$^{-1}$ kpc$^{-1}$ is also inferred
from an analysis of the motions of young open clusters  with ages
less than 100 Myr \citep{melnik2016} and Galactic maser sources
\citep{reid2009a,bobylev2012}.  However, Hipparcos proper motions
of classical Cepheids yield a slightly smaller value of
$\Omega_0$, $28.8\pm0.8$ \citep{melnik2015}.

\begin{figure}
\resizebox{\hsize}{!}{\includegraphics{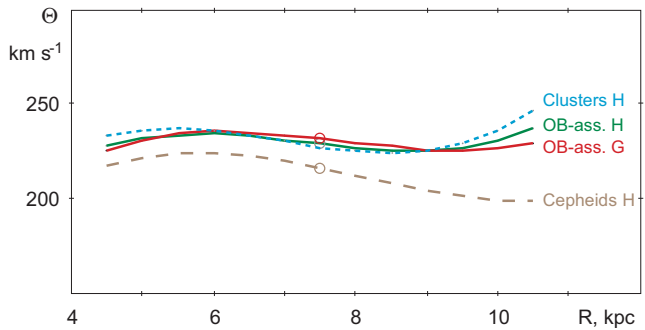}}
\caption{Galactic rotation curve obtained for different young
objects: OB-associations with TGAS (sample 1) and  Hipparcos data,
young open clusters and classical Cepheids. The rotation curves
based on TGAS and Hipparcos proper motions are labeled by the "G"
and "H" letters, respectively. The circle shows the position of
the Sun.  We can see that the rotation curves obtained for
OB-associations with  TGAS and Hipparcos proper motions agree well
with each other.} \label{rot_curve}
\end{figure}

Figure~\ref{rot_curve} shows the Galactic rotation curve obtained
for different young objects:  OB-associations with TGAS (sample 1)
and Hipparcos  \citep{melnikdambis2009} data, young open clusters
\citep{melnik2016} and classical Cepheids \citep{melnik2015}. We
can see that the rotation curves calculated for OB-associations
with  TGAS and Hipparcos proper motions are in good agreement.

%--------------------------- Table 3 ----------------------------
\begin{table*}
\caption{Parameters of the rotation curve}
 \begin{tabular}{lcccccccc}
  \hline
  Sample & $\Omega_0$ & $\Omega'_0$ & $\Omega''_0$ & $u_0$ & $v_0$  & A & $\sigma_0$ & $N_{eq}$  \\
   &  km s$^{-1}$  & km s$^{-1}$  & km s$^{-1}$  &
  km s$^{-1}$ & km s$^{-1}$  &km s$^{-1}$ &km s$^{-1}$   \\
    & kpc$^{-1}$ & kpc$^{-2}$ & kpc$^{-3}$ &  &  & kpc$^{-1}$  & \\
  \\[-7pt] \hline\\[-7pt]
1 ($n_{vr}\ge 2$, $n_{\mu}\ge2$) & 31.08 & -4.75 & 1.11 & 7.27 & 10.63 & 17.83 & 7.42 & 126 \\
   & $\pm0.86$ & $\pm0.18$ & $\pm0.22$ & $\pm1.01$ & $\pm1.33$ &  $\pm0.68$ &\\
 \hline
2 ($n_{vr}\ge 5$, $n_{\mu}\ge5$) & 31.31 & -4.68 & 1.17 & 7.16 & 12.68 & 17.54 & 7.69 & 82\\
   & $\pm1.06$ & $\pm0.22$ & $\pm0.31$ & $\pm1.28$ & $\pm1.77$ &  $\pm0.86$ & & \\
 \hline
\label{tab_rot_curve}
\end{tabular}
\end{table*}
%------------------------------------------------------------------------------

%-----------------------    Figure 2   ----------------------------------------
%------------------------------------------------------------------------------

\subsection{Residual velocities}

The residual velocities are determined as the differences between
the observed heliocentric velocities and  the velocities due to
the inferred rotation curve and the solar motion toward the apex
($V_{res}=V_{obs} - V_{rot}-V_{ap}$). The residual velocities
characterize non-circular motions in the Galactic disc. We
consider the residual velocities in the direction of the Galactic
radius-vector, $V_R$, and in the azimuthal direction, $V_T$. The
velocity components $V_R$ and $V_T$ are positive when directed
away from the Galactic centre and in the direction of the Galactic
rotation, correspondingly.

%-----------------------    Figure 3   ----------------------------------------
\begin{figure}
\resizebox{\hsize}{!}{\includegraphics{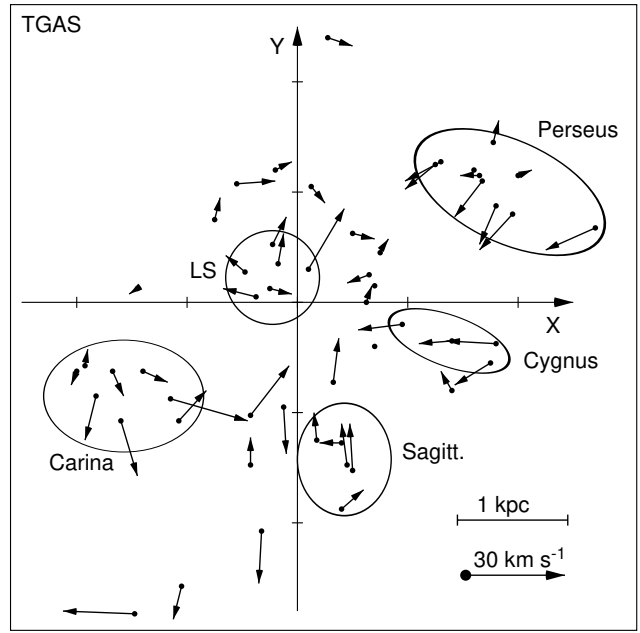}}
\caption{Distribution of the residual velocities of
OB-associations  in the Galactic plane. Shown are the residual
velocities  derived with kinematical data of at least two stars
with known line-of-sight velocities and TGAS proper motions.
OB-associations with residual velocities $|V_R|$ and $|V_T|$
smaller than 3 km s$^{-1}$ are shown as the black circles without
any vector. The ellipses indicate the positions of the
Sagittarius, Scorpio, Carina, Cygnus, Local System (LS), and
Perseus stellar-gas complexes. The X-axis points in the direction
of Galactic rotation and the Y-axis is directed away from the
Galactic centre. One tick interval along the X- and Y-axis
corresponds to 1 kpc. The Sun is at the origin.} \label{res_vel}
\end{figure}
%------------------------------------------------------------------------------
Figure~\ref{res_vel} shows the distribution of residual velocities
of OB-associations in the Galactic plane derived with TGAS proper
motions. It also  shows the positions of the Sagittarius, Scorpio,
Carina, Cygnus, Local System, and Perseus stellar-gas complexes
from the list by \citet{efremov1988}. We can see that in some
complexes the residual velocities $V_R$ and $V_T$ have a preferred
direction. In the Perseus complex the average residual velocity
components are $V_R=-5.5\pm2.3$ and $V_T=-4.7\pm1.6$, whereas in
the Sagittarius complex they are equal to $V_R=+7.9\pm2.5$ and
$V_T=-1.0\pm2.1$ km s$^{-1}$.  The distribution of residual
velocities can be easily explained in terms of a model of the
Galaxy with a two-component outer ring $R_1R_2$
\citep{melnikrautiainen2009,rautiainen2010,melnik2011}. The
residual velocities in other complexes are: $V_R=+6.9\pm2.7$ and
$V_T=1.2\pm3.1$ (Local System), $V_R=-5.4\pm2.9$ and
$V_T=+4.3\pm3.1$ (Carina) and $V_R=-4.6\pm1.5$ and
$V_T=-11.5\pm1.4$ km s$^{-1}$ (Cygnus). The rms difference of
residual velocities in the stellar-gas complexes derived from TGAS
and Hipparcos proper motions is less than 1 km s$^{-1}$.

Currently, the accuracy of measured line-of-sight velocities of
stars in OB-associations   is less than the accuracy of sky-plane
velocities derived with TGAS proper motions. The new epoch in the
study of Galactic kinematics will begin when more accurate
line-of-sight velocities are obtained for young stars by the {\it
Gaia} radial-velocity spectrometer \citep{gaia2016a}. Of
particular interest is the direction toward the Sagittarius arm
\citep{grosbol2016}.

\subsection{Motions in the z-direction}

The residual velocities in the z-direction are determined from the
proper motions $\mu_b$ and line-of-sight velocities $V_r$:

\begin{equation}
V_z = 4.74 \mu_b \cos b \cdot  r + V_r \sin b + w_0,
\end{equation}

\noindent where $w_0$ is the velocity of the Sun  along the
Z-axis. Factor 4.74 transforms proper motions determined in units
of mas yr$^{-1}$ into km s$^{-1}$ provided $r$ is  in kpc.

We selected 53 OB-associations containing at least two stars with
both known line-of-sight velocities and TGAS proper motions to
determine their  $V_z$ velocities and to calculate the most
probable value of the solar velocity $w_0$, which appears to be
$w_0=7.67\pm0.52$ km s$^{-1}$. The rms deviation of velocities in
the z-direction computed with TGAS catalog data is
$\sigma_{vz}=3.80$ km s$^{-1}$, which is slightly less than the
scatter of vertical velocity components computed using  Hipparcos
data, $\sigma_{vz}=5.0$ km s$^{-1}$.

Note that the  $V_z$ velocities of OB-associations R~103 ($r=3.2$
kpc) and Ara OB1B (2.8 kpc) decrease in absolute value nearly
twice: from $V_z=-31$ and -24 km s$^{-1}$ when computed with
Hipparcos proper motions to $V_z=-15$ and -8 km s$^{-1}$ if
determined using TGAS data. The other association with large $V_z$
velocity is Cyg OB8, $V_z=+11$ km s$^{-1}$.  Excluding these three
associations changes the value of $w_0$ to $7.45\pm0.38$ and
decreases $\sigma_{vz}$ to 2.69 km s$^{-1}$.

Interestingly, the $V_z$ velocity of OB-association Per OB1 is
nearly zero $V_z=-0.9$ km s$^{-1}$ and though Per OB1 is located
$\sim 100$ pc below the Galactic plane, $b=-3.15^\circ$, it  does
not  move  practically as a whole along the Z-axis.

\section{Internal properties of OB-associations}

\subsection{Velocity dispersion in OB-associations with TGAS proper motions}

%-----------------------    Figure 4   ----------------------------------------
\begin{figure}
\resizebox{\hsize}{!}{\includegraphics{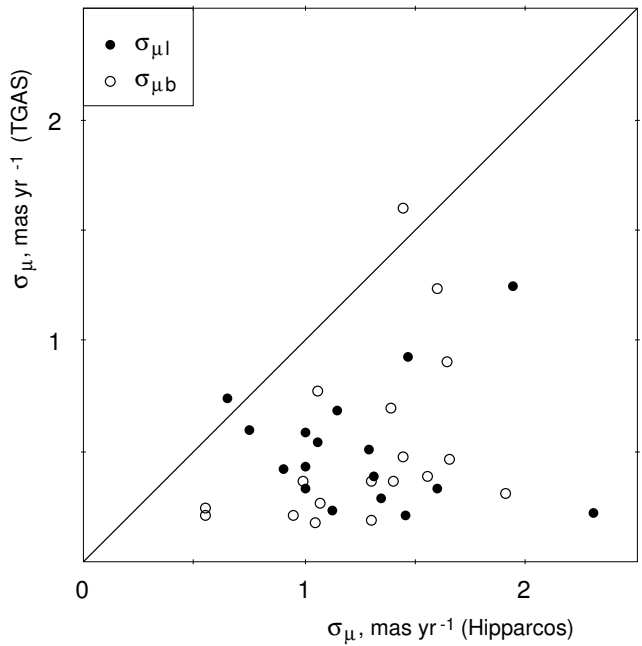}}
\caption{Comparison of the dispersions of  proper motions inside
OB-associations calculated with  Hipparcos and TGAS data. Velocity
dispersions along the l- and b-coordinate, $\sigma_{\mu l}$ and
$\sigma_{\mu b}$, are shown by the black and white circles,
respectively.  TGAS dispersions can be seen, on average, to be
smaller than the corresponding Hipparcos dispersions. The
$\sigma_{\mu l}=3.8$ mas yr$^{-1}$ of association Cyg OB7, nearly
the same when computed with TGAS and Hipparcos proper motions, is
excluded from consideration.} \label{disper_com}
\end{figure}
%------------------------------------------------------------------------------

We compare  the dispersions of  proper motions of stars in
OB-associations obtained with  Hipparcos and TGAS data
(Fig.~\ref{disper_com}).  Here $\sigma_{\mu l}$ and $\sigma_{\mu
b}$ are the robust estimates of the dispersions of proper motions
in OB-associations derived in the following way. First, we sort
the  proper motions $\mu_l$ or $\mu_b$ of TGAS stars in
OB-association to make rows of increasing values. Second, we
exclude equal number of objects constituting the (1-0.68)/2
fraction of  stars, $n_\mu$, from the upper and lower sides of the
distribution.  The remaining stars make up for nearly 68 percent
of the  number $n_\mu$. The half-width of the interval in the
proper-motion components spanned by the remaining stars amounts to
$\sigma_{\mu l}$ or $\sigma_{\mu b}$, respectively. It is an
analog of the root-mean-squared velocity dispersion but it is less
sensitive to outliers. We consider 18 OB-associations containing
at least 10 stars with TGAS proper motions. When computed with
TGAS data, the proper motion dispersions along l- and
b-coordinates, $\sigma_{\mu l}$ and $\sigma_{\mu b}$, decrease
significantly compared to the estimates obtained with  Hipparcos
proper motions: the reduction factor is $0.37\pm0.006$ and
$0.40\pm0.006$ for l- and b-directions, respectively, or
$0.39\pm0.04$ if averaged over both components. Note that
$\sigma_{\mu l}$ of OB-association Cyg OB7 was excluded from
consideration: it is very large, $\sigma_{\mu l}=3.8$ mas
yr$^{-1}$, and it is nearly the same when computed with TGAS and
Hipparcos data. However, because of the small heliocentric
distance of Cyg OB7, $r=0.63$ kpc, even this very large
$\sigma_{\mu l}$ value translates into a velocity dispersion of
only $\sigma_{vl}=11.4$ km s$^{-1}$.

We calculate the  TGAS velocity dispersions, $\sigma_{vl}$ and
$\sigma_{vb}$,  along l- and b-coordinates as follows:

\begin{equation}
 \sigma_{vl} = 4.74r\sigma_{\mu l},
\end{equation}
\begin{equation}
 \sigma_{vb} = 4.74r\sigma_{\mu b},
\end{equation}

Table~\ref{mass} lists the  TGAS velocity dispersions,
$\sigma_{vl}$ and  $\sigma_{vb}$, obtained for 18 OB-associations
containing at least 10 stars with TGAS proper motions, $n_\mu \ge
10$. It also provides the general information for OB-associations: the
average Galactic coordinates, $l$ and $b$,  heliocentric ($r$) and
Galactocentric distance ($R$) distances.

The average  TGAS velocity dispersions in
OB-associations are $\overline{\sigma_l}=4.3$ and
$\overline{\sigma_b}=3.4$ km s$^{-1}$. It is currently not clear
why the velocity dispersions in Cyg OB8 ($\sigma_b=13.9$) and
in Cyg OB7 ($\sigma_l=11.4$) are so high.

Generally, the Cygnus region  is difficult to analyze because of
the high concentration of bright stars in the sky plane. The large
velocity dispersions in associations Cyg OB7 ($\sigma_{vl}$) and
Cyg OB8 ($\sigma_{vb}$) compared to other associations  can also
be seen in Figure~\ref{expan_all_1}.

The fact that the TGAS velocity dispersions inside OB-associations
drops to  4 km s$^{-1}$, whereas  the standard deviation,
$\sigma_0$, of  the velocities of OB-associations  from the
rotation curve remains at the level 7--8 km s$^{-1}$
(Table~\ref{tab_rot_curve}) indicates that OB-associations
identified by \citet{blahahumphreys1989} mainly include stars born
from the same molecular cloud.

Due to the turbulent motions inside giant molecular clouds, the
velocity dispersion inside a cloud must grow with  the size $S$ of
the region considered as $\sigma_v \sim S^p$. The study by
\citet{solomon1987} suggests that power index $p$ is close to
$p\approx0.5$. If we suppose that the velocity dispersion in a
giant cloud with a diameter of 10 pc (the average size of open
clusters) is 1--2 km s$^{-1}$, then it must be $\sim$3 times
larger for clouds of 100 pc in diameter (the average size of
OB-associations), i.e. 3--6 km s$^{-1}$. So the value of
$\sigma_v=4$ km s$^{-1}$ is consistent with the properties of
giant molecular clouds.

\subsection{Binary stars}

%-----------------------    Figure 5   ----------------------------------------
\begin{figure}
\resizebox{\hsize}{!}{\includegraphics{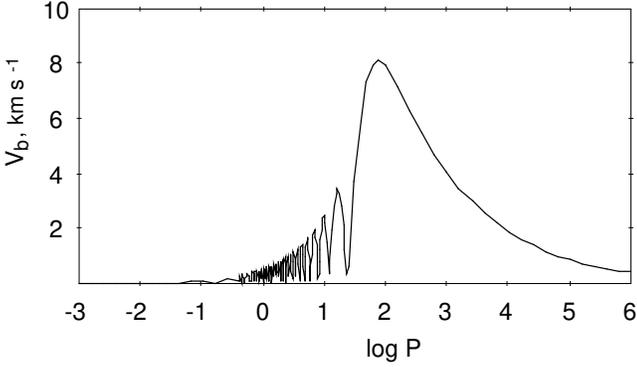}}
\caption{Dependence of the additional velocity $V_b$ on $\log P$.
Binary system considered includes two stars of mass 10M$_\odot$
rotating on a circular orbit around their common mass center with
period $P$. The velocity $V_b$ is the maximal contribution of a
binary system into the TGAS velocity dispersion. }
\label{binary_stars}
\end{figure}
%------------------------------------------------------------------------------

\citet{mason1998} find that the majority ($>59$ percent) of O
stars in clusters and OB-associations are binary. In principle,
binary stars can  inflate the observed velocity dispersion  inside
OB-associations derived with TGAS proper motions, which are
calculated from the positions obtained with the time difference
$T=24$ yr. The main contribution into TGAS proper motions must be
provided by  binary systems with orbital periods of $P\approx50$
yr, which corresponds to the component of  the binary system
moving to the  opposite points of their orbit between  the epochs
of Hipparcos and {\it Gaia}.

Let us calculate the upper limit for the probable  contribution of
binary stars, $\sigma_b$, into the observed dispersion of TGAS
proper motions inside OB-associations. For simplicity we consider
a binary system consisting of two components  of mass
$M_b=10$M$_\odot$ moving in circular orbits around their common
mass center with period $P$. A 10M$_\odot$-star can be considered
as a representative of stars in OB-associations from the list by
\citet{blahahumphreys1989}: most of stars listed in the catalog
(about 2/3) have spectral types in the B0--B2 interval
corresponding to the mass range 8--15$M_\odot$ \citep{hohle2010}.

The distance between the component stars $D$ can be estimated from
Newton's gravity law:

\begin{equation}
\frac{M_b \cdot V_{orb}^2}{D/2}=\frac{G \cdot M_b^2}{D^2},
\end{equation}

\noindent or, given that

\begin{equation}
V_{orb} = \pi \cdot D/P,
\end{equation}

\noindent we have

\begin{equation}
\frac{D^3}{P^2}=\frac{G \cdot M_b}{2\pi^2},
\end{equation}

\noindent or

\begin{equation} \label{binsep}
D = (\frac{G}{2\pi^2})^{1/3} \cdot P^{2/3} \cdot M_b^{1/3} ,
\end{equation}

\noindent The distance between components $D$ is the maximal
possible shift of the star between the two  epochs of
observations. However, the real  displacement is smaller because
the star positions at the two epochs are practically always less
than the orbit diameter apart. The length of the chord connecting
the initial and final position of the star in its circular orbit
is   (by law of cosines):

\begin{equation} \label{smax2}
S =  D \cdot \sqrt{\frac{(1 -  cos (2 \pi \frac{T}{P}))}{2}}.
\end{equation}

The maximum possible extra velocity along any given direction due
to the orbital displacement of this component is equal to

\begin{equation}
V_b = \frac{S}{T},
\end{equation}

\noindent or, in view of eqs.~(\ref{binsep}) and (\ref{smax2}):

\begin{equation}
V_b = 6.0  \cdot P^{2/3} \cdot M_b^{1/3} T^{-1} \cdot
\sqrt{\frac{(1 -  cos (2 \pi \frac{T}{P}))}{2}}
\end{equation}

\noindent    where $V_b$ is  in km~s$^{-1}$, $M_b$ is in solar
masses,  $P$ and $T$ are in years.

Figure~\ref{binary_stars} shows the dependence of the maximum
additional velocity $V_b$ on  $\log P$ for $M_b$~=~10$M_{\odot}$
and $T$~=~24~yr. The velocity $V_b$ reaches maximum of 8.1 km
s$^{-1}$ at period $\sim 80$ yr, because the radius of the orbit
increases with increasing $P$. The decrease of $V_b$ at larger $P$
is caused by the fact that a star can pass only a small part of
its orbit for 24-year time interval. On the other hand, the
decrease of $V_b$ at smaller $P$ is due to decreasing distance $D$
between binary stars, $D \sim P^{2/3}$. The velocity $V_b$
demonstrates oscillations with zero values corresponding to the
periods $P = T/n$, where $n$ is integer and the binary star makes
a whole number of revolutions in time interval $T$, resulting in
zero displacement $D$.

To estimate the effective contribution of binary stars into  the
velocity dispersion inside OB-association, $\sigma_v$,  we should
consider the fraction of multiple stars, $f_b$, in
OB-associations, the factor measuring the average projection,
$f_j$, of the shift $S$ into l- or b-direction,  distribution of
binary periods $f_p (\log P$), and the  additional velocity due to
orbital motion, $V_b^2 (\log P)$, as a function of log~$P$.

\citet{sana2017} overviews recent estimates of the fraction of
binary stars among OB-stars, which  vary from $f_b=0.2$ to 0.7. We
adopt $f_b=0.5$ as some average value.

If we suppose that the ensemble the binary systems in an
OB-association is oriented randomly with respect to the line of
sight then the effective value of the squared projection of the
orbital displacement $S_l^2$ or $S_b^2$ onto the longitude or
latitude direction, respectively, is equal to $S^2/3$, hence
$f_j$~=~1/3.

\citet{aldoretta2015} show that the period distribution of massive
binary stars is approximately flat  in increments of $\log P$
(\"Opik's law) in the interval of $\log P$ from -3 to +6. So we
can suppose that $f_p=1/9$ for unit of $\log P$.

Generally,  we can write:

\begin{equation}\label{sigma_b}
\sigma_b^2=   f_b \cdot f_j \cdot \int_{\log P=-3}^{\log P=+6}
V_b^2(\log P) \cdot f_p (\log P) \cdot d \log P.
\end{equation}

We now use the values of $f_b=0.5$, $f_j=1/3$, $f_p=1/9$ and the
distribution of $V_b$ over $\log P$ (Fig.~\ref{binary_stars}) to
calculate the effective contribution of binary stars into the TGAS
velocity dispersion (Eq.~\ref{sigma_b}), which appears to be
$\sigma_b=1.19$ km s$^{-1}$, or

\begin{equation}
\sigma_b=1.19 \cdot(M_b/10 M_{\odot})^{1/3}
\end{equation}

\noindent for a binary component of mass $M_b$. This value is
rather small compared to $\sigma_v=4$ km s$^{-1}$ caused by
turbulent motions inside giant molecular clouds.

\subsection{Virial  and stellar masses of OB-associations}

Giant molecular clouds are supposed to be close to their  virial
equilibrium \citep{larson1981,krumholz2006} and therefore the
velocity dispersions of stars in OB-associations must correspond
to the masses of their parent giant molecular clouds. The high
precision of TGAS proper motions  allows estimating the virial
masses of OB-associations. The virial condition between the
kinetic K and potential U energies of a system is:

\begin{equation}
 2K+U=0.
\end{equation}

The potential energy of a uniform sphere of radius $a$ and
constant density $\rho$ is :

\begin{equation}
 U=-\int\limits_0^a\frac{(\frac{4}{3}G\pi\rho r^3)(4\pi \rho r^2 dr)}{r}=-\frac{3GM^2}{5a},
\end{equation}

\noindent where M is the mass of the sphere:

\begin{equation}
 M=\frac{4\pi a^3\rho}{3}.
\end{equation}

The kinetic energy of a system with one-dimensional velocity
dispersion $\sigma_v$ is :

\begin{equation}
 K=\frac{3M\sigma_v^2}{2}.
\end{equation}

Hence, the virial mass of the parent giant molecular cloud can be
calculated in the following way:

\begin{equation}
 M_{vir}=\frac{5a\sigma_v^2}{G},
\label{m_vir}
\end{equation}

\noindent  where $\sigma_v$ and $a$  are determined from
observations.

There is no problem with observed one-dimensional velocity
$\sigma_{v}$, it can be estimated as  the average of $\sigma_{vl}$
and $\sigma_{vb}$:

\begin{equation}
 \sigma_{v}=(\sigma_{vl}+\sigma_{vb})/2.
\label{sigma_v}
\end{equation}

\noindent However, the so-called radius $a$ of a parent molecular
cloud  can be treated in different ways. We calculate $a$ as the
radius containing 68 percent of the association member stars. Here
we consider all members of OB-associations with known photometric
measurements and not just TGAS stars.  We also suppose that the
sizes of OB-associations have not changed significantly since the
epoch of star formation. However, some OB-associations appear to
be expanding, so in these cases, the radius $a$ needs a correction
(for more details see Section 4.3).

Table~\ref{mass}  also gives values of $M_{vir}$ and $a$. The
median value of $M_{vir}$ for 18 OB-associations considered is 7.1
10$^5$ M$_\odot$ which is consistent with the masses of giant
molecular clouds 10$^5$-10$^6$ M$_\odot$ \citep{sanders1985}.

Note that in our calculations of virial masses we  used the values
of $\sigma_v$ corrected for the effect  of binary stars:

\begin{equation}
 \sigma_{cor}^2=\sigma_{v}^2-\sigma_b^2.
\label{cor_sig_v}
\end{equation}

To estimate the stellar masses of OB-associations we use the
multi-component power-law distribution derived by
\citet{kroupa2002}. The number of stars $dN(M)$ in the mass range
$\Delta M$ is determined for three mass domains  by the following
laws:

\begin{equation}
 \begin{array}{cc}
 dN(M)= C_0&
 \left \{
 \begin{array}{ll}
 C_1  \cdot M^{-0.3} &  0.01 < M/M_\odot <0.08\\
 %&\\
 C_2 \cdot M^{-1.3} &  0.08 < M/M_\odot <0.5\\
  %&\\
 C_3 \cdot M^{-2.3} &  0.5 < M/M_\odot<\infty. \\
 \end{array}
 \right.\\
 \end{array}
 \label{kroupa}
\end{equation}

\noindent The coefficients $C_1$, $C_2$, and $C_3$ are calculated
from the continuity condition at the boundaries of the mass
domains: $C_1=0.469$, $C_2=0.038$ and $C_3=0.019$.

We also assume that the catalog by \citet{blahahumphreys1989}
includes all stars of OB-associations with masses greater than
$20M_\odot$ or it is essentially complete down to the absolute
magnitude $M_V<-4.0^m$. To determine the mass of stars in
OB-associations we use the mass-absolute  magnitude relation by
\citet{bressan2012}. This relation is monotonic in the range
$-7.8<M_V<+0.50$ and corresponds to the mass range 3--50
M$_\odot$, which covers 99.3 percent of OB-association stars
listed in the catalog by \citet{blahahumphreys1989}. The age of
the brightest OB-stars is supposed to be 4 Myr.

The value of $C_0$ is derived from the following calibration:

\begin{equation}
 N_{20}=C_0 C_3 (20^{-1.3}-50^{-1.3})/1.3,
\end{equation}

\noindent where $N_{20}$ is the number of stars with masses
$M>20$M$_\odot$ in a particular OB-association.

Table~\ref{mass} also gives the stellar masses $M_{st}$ of
OB-associations and  the number $N_{20}$ of stars with masses
$M>20$M$_\odot$. The median value of $M_{st}$ calculated for 18
OB-associations is 9.0 10$^3$ M$_\odot$.  Note that in the case of
pure Salpiter (1955) distribution, $dN\sim M^{-2.35}dM$, in the
mass interval from 0.08 \citep{grossman1971} to 50 M$_\odot$ we
obtain nearly twice greater   $M_{st}$ values with the median of
19 10$^3$ M$_\odot$.

For each OB-association from Table~\ref{mass} we can estimate the
average efficiency of star formation in its parent  molecular
cloud, which is equal to the ratio of the stellar mass to the gaseous mass
of the giant  cloud:

\begin{equation}
 \epsilon=M_{st}/M_{vir}.
 \label{epsilon}
\end{equation}

Table~\ref{mass} shows that $\epsilon$ values vary from 0.1 to
13.6 percent  with  the median value of 2.1 percent which agrees
with other estimates \citep{myers1986, evans2009, garcia2014}.

Note that the  estimates of star-formation efficiency $\epsilon$
obtained here refer to the entire  volume of OB-associations, and
$\epsilon$ can be smaller or larger in different parts of their
parent  giant molecular cloud.

\subsection{Present-day mass of OB-associations and their unboundness}

The fact that virial mass exceeds the stellar mass by more than 70
times supports the conclusion that OB-associations are unbound objects.
Some doubts concern the  gas  which OB-associations could harbor
inside  their volumes, so the real masses of OB-associations can
be conspicuously larger than those of the stellar component.

Numerical simulations demonstrate that stars with masses $M\ge
20$M$_\odot$ destroy crucially molecular clouds of masses up to a
few times 10$^4$ M$_\odot$ \citep{colin2013,dale2012}.

Table~\ref{mass} shows that  of the 18 OB-associations considered
nine contain more than  20 stars with masses $M\ge 20$M$_\odot$.
On the other hand, masses of their parent molecular clouds are
10$^5$--10$^6$M$_\odot$, implying the average escape velocity of
12 km s$^{-1}$, and hence the full evaporation of gas clouds seems
to be impossible.

\citet{dale2012} simulate the star-formation processes in the
molecular cloud with mass 10$^6$ M$_\odot$ and initial radius of
180 pc. They found that after 3 Myr of ionization the surface
density $\Sigma$ of the cloud is $\sim2$ 10$^{-3}$ g cm$^{-2}$,
which corresponds to the volume density of $\sim 1$ atom
cm$^{-3}$.

\citet{cappa2000} study the distribution of HI gas in the vicinity
of association Per OB1. Their Figure 2 indicates that HI surface
density inside Per OB1 is  $\Sigma\sim3$ 10$^{20}$ atoms
cm$^{-2}$. If the size of Per OB1  along the line of sight is
$\sim$200 pc then the HI volume density must be $\sim 0.5$ atom
cm$^{-3}$.

Let us suppose that the fraction of molecular gas inside the
volume of OB-associations is negligible, but neutral hydrogen with
the volume density $\rho_H$ equal to  1 atom cm$^{-3}$ can be
present there. Then the  upper estimate for the present-day mass
of OB-association, $M_t$, includes the mass of the stellar
component, $M_{st}$, and the mass of HI possibly located inside
the volume of the OB-association:

\begin{equation}
 M_t=M_{st}+M_g.
 \label{m_t}
\end{equation}

The gas mass $M_g$ can be estimated from the following formula:

\begin{equation}
 M_g=\frac{4\pi}{3} a^3 \rho_H.
 \label{m_g}
\end{equation}

Table~\ref{mass}  also lists the masses of the gaseous component, $M_g$.
The median gas mass  appears to be 18.3 10$^3$ M$_\odot$, and hence
the contribution of gas to the total mass is usually comparable to the mass
of the stellar component. The exception is two extremely extended
OB-associations: Cep OB1 and NGC 2439, where $M_g$ exceeds
$M_{st}$ by a factor of more than 30, however, even in these cases the total
masses of OB-associations, $M_t$, are nearly 10 times smaller than the
corresponding virial masses, $M_{vir}$.

We can thus  conclude that  OB-associations must be  unbound
objects provided they do not contain  a lot of dense gas.

%--------------------------- Table 4 ------------------------------------------
\begin{table*}
\caption{Virial and stellar masses of OB-associations, $M_{vir}$
and $M_{st}$, star formation efficiency $\epsilon$}
 \begin{tabular}{lrrrrrrrrrrrrrr}
 \\[-7pt] \hline\\[-7pt]
Name & $l\;$ & $b\;$ & $r\;$ & $R\;$ & $\sigma_{vl} \quad$ &
$\sigma_{vb} \quad$ &  $a\;$ & $r_{td}$ & $M_{vir}\;$ &$n_\mu$
&  $M_{st}\quad$ & $N_{20}$ & $M_g\quad$ & $\epsilon$ 100\% \\
[2pt]
& deg. & deg. & kpc & kpc &   km s$^{-1}$   &  km s$^{-1}$ & pc & pc &  $M_\odot\;$ &  & $M_\odot\quad$ & & $M_\odot\quad$ &   \\
  \\[-7pt] \hline\\[-7pt]
SGR OB1   &  7.55 & -0.77 &  1.26 &  6.25 & 1.3 & 4.6 & 42 & 24 &     3.6 $10^5$ & 13 &    9.6 $10^3$ &  23 &    7.6 $10^3$ &  2.7$\;\;$  \\
CYG OB3   & 72.77 &  2.03 &  1.83 &  7.17 & 2.6 & 1.5 & 26 & 28 &     0.9 $10^5$ & 16 &   11.7 $10^3$ &  28 &    1.7 $10^3$ & 13.6$\;\;$  \\
CYG OB1   & 75.84 &  1.12 &  1.46 &  7.28 & 3.8 & 2.5 & 32 & 31 &     3.1 $10^5$ & 12 &   16.7 $10^3$ &  40 &    3.5 $10^3$ &  5.3$\;\;$  \\
CYG OB8   & 77.91 &  3.36 &  1.83 &  7.34 & 3.7 &13.9 & 42 & 24 &    36.8 $10^5$ & 10 &    7.1 $10^3$ &  17 &    7.4 $10^3$ &  0.2$\;\;$  \\
CYG OB7   & 88.99 &  0.03 &  0.63 &  7.52 &11.3 & 2.0 & 53 & 19 &    26.4 $10^5$ & 16 &    3.3 $10^3$ &   8 &   15.6 $10^3$ &  0.1$\;\;$  \\
CEP OB2   &101.59 &  4.64 &  0.73 &  7.68 & 3.2 & 4.2 & 45 & 26 &     6.5 $10^5$ & 34 &    8.3 $10^3$ &  20 &    9.4 $10^3$ &  1.3$\;\;$  \\
CEP OB1   &104.14 & -0.97 &  2.78 &  8.61 & 9.0 & 4.7 &178 & 36 &    94.8 $10^5$ & 20 &   16.7 $10^3$ &  40 &  576.4 $10^3$ &  0.2$\;\;$  \\
CEP OB3   &110.42 &  2.56 &  0.70 &  7.77 & 1.7 & 1.4 & 12 & 14 &     0.1 $10^5$ & 13 &    1.2 $10^3$ &   3 &    0.2 $10^3$ & 10.0$\;\;$  \\
PER OB1   &134.67 & -3.15 &  1.83 &  8.88 & 5.0 & 3.1 & 59 & 48 &     8.2 $10^5$ & 58 &   36.2 $10^3$ &  87 &   20.9 $10^3$ &  4.4*  \\
CAS OB6   &135.02 &  0.75 &  1.75 &  8.83 & 2.4 & 3.9 & 78 & 29 &     7.8 $10^5$ & 11 &    7.9 $10^3$ &  19 &   49.4 $10^3$ &  1.0$\;\;$  \\
CAM OB1   &141.08 &  0.89 &  0.80 &  8.14 & 4.7 & 3.4 & 86 & 23 &    15.3 $10^5$ & 26 &    5.0 $10^3$ &  12 &   66.0 $10^3$ &  0.3$\;\;$  \\
AUR OB1   &173.83 &  0.14 &  1.06 &  8.55 & 2.6 & 1.5 & 69 & 22 &     2.2 $10^5$ & 12 &    3.7 $10^3$ &   9 &   33.9 $10^3$ &  1.7$\;\;$  \\
MON OB2   &207.46 & -1.65 &  1.21 &  8.59 & 1.3 & 2.6 & 71 & 25 &     2.0 $10^5$ & 10 &    5.8 $10^3$ &  14 &   36.6 $10^3$ &  2.9$\;\;$  \\
NGC 2439  &245.27 & -4.08 &  3.50 &  9.50 & 7.2 & 3.0 &155 & 31 &    44.0 $10^5$ & 10 &    7.9 $10^3$ &  19 &  384.7 $10^3$ &  0.2$\;\;$  \\
CAR OB1   &286.45 & -0.46 &  2.01 &  7.19 & 7.0 & 2.3 & 63 & 34 &    11.9 $10^5$ & 15 &   21.2 $10^3$ &  51 &   26.2 $10^3$ &  1.8*  \\
CAR OB2   &290.39 &  0.12 &  1.79 &  7.08 & 2.0 & 2.2 & 28 & 26 &     1.0 $10^5$ & 10 &   10.4 $10^3$ &  25 &    2.2 $10^3$ & 10.5$\;\;$  \\
CRU OB1   &294.87 & -1.06 &  2.01 &  6.90 & 3.1 & 2.0 & 40 & 26 &     2.4 $10^5$ & 17 &   10.0 $10^3$ &  24 &    6.8 $10^3$ &  4.2$\;\;$  \\
CEN OB1   &304.14 &  1.44 &  1.92 &  6.62 & 5.2 & 1.9 & 69 & 32 &     9.1 $10^5$ & 28 &   21.2 $10^3$ &  51 &   34.4 $10^3$ &  2.3$\;\;$  \\
    \\[-7pt] \hline\\[-7pt]
\multicolumn{13}{l}{* Values of  $M_{vir}$ and  $\epsilon$ for Per
OB1 and Car OB1 are corrected for the expansion effect (Section
4.3)}
  \end{tabular}
 \label{mass}
\end{table*}
%----------------------------------------------------------------------------------

%--------------------------- Table 5 ----------------------------------------------
\begin{table*}
\caption{Expansion/compression  of OB-associations with TGAS
proper motions}
 \begin{tabular}{lrrrrrrrrr}
 \\[-7pt] \hline\\[-7pt]
Name &    $p_l \quad \quad$ & $p_b \quad \quad$ & $a\;$  & $e_1\quad \;$& $u_l\quad \quad $&$u_b\quad \quad$& $\tilde u_l\quad \quad $&$\tilde u_b\quad \quad$ & $n_\mu$ \\
[2pt]
&  km s$^{-1}$ kpc$^{-1}$ & km s $^{-1}$ kpc$^{-1}$ & pc  & km s$^{-1}$ & km s$^{-1}$ $\;$& km s$^{-1}$ $\;$ & km s$^{-1}$ $\;$& km s$^{-1}$ $\;$ &\\
  \\[-7pt] \hline\\[-7pt]
SGR OB1   & $  70\pm 38$ & $ 238\pm  26$ & 42 &  $ 0.3\pm 0.1$ &  $  2.9\pm 1.6$ & $ 10.0\pm 1.1$ & $  2.6\pm 1.6$ & $\underline{ 9.7\pm 1.1}$ & 13 \\
CYG OB3   & $ 139\pm138$ & $ 159\pm 116$ & 26 &  $ 0.1\pm 0.0$ &  $  3.6\pm 3.6$ & $  4.1\pm 3.0$ & $  3.5\pm 3.6$ & $  4.0\pm 3.0$ & 15 \\
CYG OB1   & $ -57\pm102$ & $ 112\pm 131$ & 32 &  $ 0.3\pm 0.1$ &  $ -1.8\pm 3.3$ & $  3.6\pm 4.2$ & $ -2.1\pm 3.3$ & $  3.3\pm 4.2$ & 12 \\
CYG OB8   & $ -45\pm 55$ & $ 186\pm 152$ & 42 &  $ 0.5\pm 0.1$ &  $ -1.9\pm 2.3$ & $  7.7\pm 6.3$ & $ -2.4\pm 2.3$ & $  7.3\pm 6.3$ & 10 \\
CYG OB7   & $ -39\pm 85$ & $  63\pm  26$ & 53 &  $ 0.8\pm 0.2$ &  $ -2.1\pm 4.5$ & $  3.4\pm 1.4$ & $ -2.9\pm 4.5$ & $  2.6\pm 1.4$ & 16 \\
CEP OB2   & $  77\pm 26$ & $  11\pm  26$ & 45 &  $ 1.1\pm 0.1$ &  $  3.5\pm 1.2$ & $  0.5\pm 1.2$ & $  2.4\pm 1.2$ & $ -0.6\pm 1.2$ & 34 \\
CEP OB1   & $  56\pm 19$ & $   1\pm  39$ &178 &  $ 3.7\pm 0.1$ &  $  9.9\pm 3.4$ & $  0.1\pm 6.9$ & $  6.2\pm 3.4$ & $ -3.6\pm 6.9$ & 20 \\
CEP OB3   & $ -39\pm 48$ & $  71\pm  67$ & 12 &  $ 0.4\pm 0.0$ &  $ -0.5\pm 0.6$ & $  0.9\pm 0.8$ & $ -0.9\pm 0.6$ & $  0.5\pm 0.8$ & 12 \\
PER OB1   & $  45\pm 15$ & $ 103\pm  22$ & 59 &  $ 1.4\pm 0.0$ &  $  2.7\pm 0.9$ & $  6.1\pm 1.3$ & $  1.3\pm 0.9$ & $\underline{4.7\pm 1.3}$ & 58 \\
CAS OB6   & $   9\pm 47$ & $ 227\pm  80$ & 78 &  $ 1.9\pm 0.1$ &  $  0.7\pm 3.7$ & $ 17.8\pm 6.3$ & $ -1.2\pm 3.7$ & $ 15.9\pm 6.3$ & 11 \\
CAM OB1   & $  34\pm 20$ & $   3\pm  28$ & 86 &  $ 1.2\pm 0.2$ &  $  2.9\pm 1.8$ & $  0.3\pm 2.4$ & $  1.7\pm 1.8$ & $ -0.9\pm 2.4$ & 26 \\
AUR OB1   & $ -36\pm 27$ & $  19\pm  10$ & 69 &  $ 0.1\pm 0.3$ &  $ -2.5\pm 1.9$ & $  1.3\pm 0.7$ & $ -2.6\pm 1.9$ & $  1.2\pm 0.7$ & 12 \\
MON OB2   & $-278\pm 77$ & $ 232\pm 203$ & 71 &  $-1.3\pm 0.2$ &  $-19.7\pm 5.4$ & $ 16.5\pm14.4$ & $\underline{-18.4\pm 5.4}$ & $ 17.8\pm14.4$ & 10 \\
NGC 2439  & $  64\pm 42$ & $-106\pm  21$ &155 &  $-2.8\pm 0.0$ &  $ 10.0\pm 6.6$ & $-16.4\pm 3.2$ & $ 12.8\pm 6.6$ & $\underline{-13.6\pm 3.2}$ & 10 \\
CAR OB1   & $ 118\pm 43$ & $ 111\pm  28$ & 63 &  $ 0.2\pm 0.1$ &  $  7.5\pm 2.7$ & $  7.0\pm 1.8$ & $\underline{  7.3\pm 2.7}$ & $\underline{  6.8\pm 1.8}$ & 15 \\
CAR OB2   & $-100\pm136$ & $  42\pm  64$ & 28 &  $ 0.1\pm 0.0$ &  $ -2.8\pm 3.8$ & $  1.2\pm 1.8$ & $ -2.9\pm 3.8$ & $  1.1\pm 1.8$ & 10 \\
CRU OB1   & $ -29\pm 47$ & $  29\pm  23$ & 40 &  $ 0.1\pm 0.0$ &  $ -1.2\pm 1.9$ & $  1.2\pm 0.9$ & $ -1.3\pm 1.9$ & $  1.1\pm 0.9$ & 17 \\
CEN OB1   & $ -87\pm 33$ & $  35\pm  18$ & 69 &  $ 0.7\pm 0.1$ &  $ -6.0\pm 2.3$ & $  2.4\pm 1.3$ & $\underline{ -6.7\pm 2.3}$ & $  1.7\pm 1.3$ & 27 \\
\hline \end{tabular} \label{par_expansion}
\end{table*}
%----------------------------------------------------------------------------------

\subsection{Tidal radius  of OB-associations}

Given the known present-day masses of OB-associations, we can
estimate their tidal radii, $r_{td}$, at which the gravitational
force produced by the OB-association equals the tidal perturbation
from the Galaxy.

Let us consider the forces applied to  a test particle located
between the OB-association and  the Galactic centre. Subscripts 1
and 2 refer to the distances and velocities of the test particle
and the centre of the OB-association, respectively. The distance
between the test particle and the centre of the OB-association is
$\Delta R=R_1-R_2<0$. The test particle is subject to two forces
working in the opposite directions: the gravity from the Galaxy,
$\Omega_1^2 R_1$, and the gravity from the OB-association, $G
M/(\Delta R)^2$. If the test particle is bound with the
OB-association, then it rotates around  the Galactic centre at the
angular velocity of the OB-association, $\Omega_2$. Here we ignore
the velocity of the test particle due to the gravity of the
OB-association. The acceleration of the test particle is equal to
the sum of  its centripetal acceleration, $\Omega_2^2 R_1$, and
additional acceleration determined with respect to the centre of
the OB-association, $\ddot R_1$. Newton's law gives us the
following equation:

\begin{equation}
 \ddot R_1-\Omega_2^2 R_1 = \frac{G M}{(\Delta R)^2}-\Omega_1^2 R_1,
 \label{tidal_1}
\end{equation}

\noindent  where $M$ is the mass of the OB-association, $M_t$. The
tidal radius, $r_{td}$ is derived  from  the condition $\ddot
R_1=0$. By assuming $\Delta R/R_2\ll 1$, we obtain the  following
estimate:

\begin{equation}
r_{td}=|\Delta R|= \left(\frac{GM}{4A\Omega}\right)^{1/3},
 \label{tidal_2}
\end{equation}

\noindent where $A$  is Oort's constant $A=-0.5R\Omega'(R)$
at the Galactocentric distance $R$.

The formula for the tidal force, $T=-4A\Omega\Delta R$, can also be
derived from basic equations by \citet[][Eq.
32]{goldreich1980},  it is also discussed  in \citet[][Eq.
1]{stark1978}.

We calculate the tidal radii $r_{td}$ for 18 OB-associations
containing at least 10 stars with TGAS proper motions.  The
angular velocity $\Omega$ and its first derivative $\Omega'$ at
the Galactocentric radius $R$ of each OB-association are computed
based on the parameters of the rotation curve given in
Table~\ref{tab_rot_curve}. The median tidal radius $r_{td}$
appears to be  40 pc, which  is a bit smaller than the median
radius $a$ of OB-associations, 56 pc. On average, 27 percent of
stars of OB-associations must lie outside their tidal radii and
tidal perturbations from the Galaxy are essential for them.

We also calculate the minimal value of the tidal radius
(Table~\ref{mass}) assuming that the mass of the OB-association
consists of stellar component exclusively, $M_t=M_{st}$. The tidal
radii of OB-associations then drop to the values in the 14--36 pc
interval with the median of 26 pc.  It means that 39 percent of
member stars of OB-associations must be located outside their
tidal radii.

Thus, two different estimates of tidal radii of OB-associations
yield nearly the same result:  $\sim 1/3$ of stars of
OB-associations must lie outside their tidal radius.

\section{Expansion of OB-associations}

\subsection{Determination of the parameters of  expansion/compression  of OB-associations}

We use TGAS proper motions to study  possible
expansion/compression of OB-associations in the l- and
b-direction. The parameters of expansion/compression $p_l$ and
$p_b$ are calculated from the following equations:

\begin{equation}
 4.74  \mu_l  r = v_{l0}+ p_l \cdot r \sin(l-l_0),
\end{equation}
\begin{equation}
 4.74  \mu_b  r = v_{b0}+ p_b \cdot r \sin(b-b_0),
\end{equation}

\noindent where $v_{l0}$ and $v_{b0}$ are the average velocities
of the association; $l_0$ and $b_0$ are the coordinates of the
centre of the association, and parameters $p_l$ or $p_b$
characterize expansion/compression along the l- or b-direction
(positive and negative values correspond to expansion and
compression, respectively).

The observed specific velocities of expansion or compression,
$u_l$ and $u_b$, are calculated in following way:

\begin{equation}
 u_l = p_l \cdot a,
\end{equation}
\begin{equation}
u_b =  p_b \cdot a.
\end{equation}

Note that  expansion/compression of OB-associations requires some
caution in interpretation:  the motion of an association as a
whole with the line-of-sight velocity $V_r$ (see Table 1) can
cause the effect of spurious expansion/compression. The velocity
of spurious expansion/compression, $e_1$,  caused by the
line-of-sight velocity $V_r$ is determined by the following
expression:

\begin{equation}
 e_1 = - V_r \cdot \frac{a}{r},
 \label{e1}
\end{equation}

\noindent where the negative line-of-sight velocities $V_r$
produces spurious expansion, $e_1>0$, while positive $V_r$ gives
spurious compression, $e_1<0$.

Table~\ref{par_expansion}   lists the parameters of observed
expansion/compression, $p_l$ and $p_b$, the radius  of
OB-associations, $a$,  the velocity of spurious
expansion/compression $e_1$, the observed specific velocities of
expansion/compression, $u_l$ and $u_b$,  and the number of stars
with known TGAS proper motions, $n_{\mu}$. We can see that  the
velocity of spurious expansion/compression, $e_1$, can reach $\sim
4$ km s$^{-1}$ (Cep OB1). So the observed velocities of
expansion/compression, $u_l$ or $u_b$, should be corrected for
this effect:

\begin{equation}
\tilde u_l = u_l - e1,
\end{equation}
\begin{equation}
\tilde u_b =  u_b - e1.
\end{equation}

\noindent The corrected velocities of expansion/compression,
$\tilde u_l$ and $\tilde u_b$, are  listed in
Table~\ref{par_expansion} as well. Further, we will discuss only
velocities, $\tilde u_l$ and $\tilde u_b$, determined at
$P>2.5\sigma$ confidence level, which are underlined in
Table~\ref{par_expansion}.

%-----------------------    Figure 6   ----------------------------------------
\begin{figure*}
\resizebox{\hsize}{!}{\includegraphics{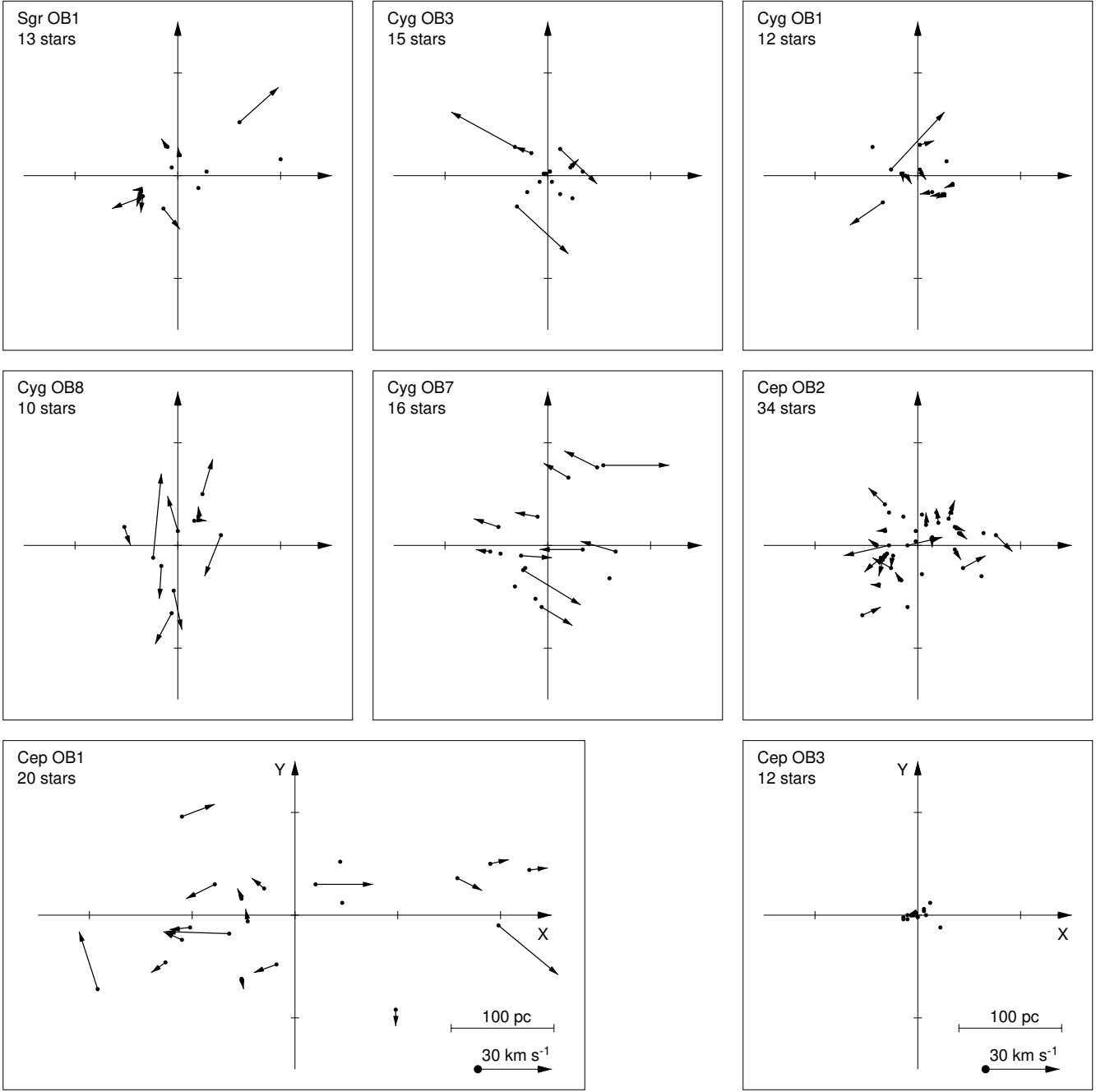}} \caption{
Distribution of the relative velocities $v_l'$ and $v_b'$
(Eq.~\ref{vl'} and ~\ref{vb'})  inside OB-associations with more
than 10  TGAS stars. Stars with the relative velocities $|v_l'|$
and $|v_b'|$ smaller than 3 km s$^{-1}$ are shown as black circles
without any vector. The axes $X$ and $Y$ are directed toward
increasing values  of Galactic coordinates $l$ and $b$,
respectively. All images are on the same scale. One tick
corresponds to 100 pc. The Per OB1 association is not included in
this panorama -- we show it in a special image
(Fig.~\ref{per_ob1}). The Cep OB1 association is shown in a wider
box because of its large extension in the l-direction. The
velocity scale is shown in the bottom row.} \label{expan_all_1}
\end{figure*}
%------------------------------------------------------------------------------
%\addtocounter{figure}{-1}
%-----------------------    Figure 7   ----------------------------------------
\begin{figure*}
\resizebox{\hsize}{!}{\includegraphics{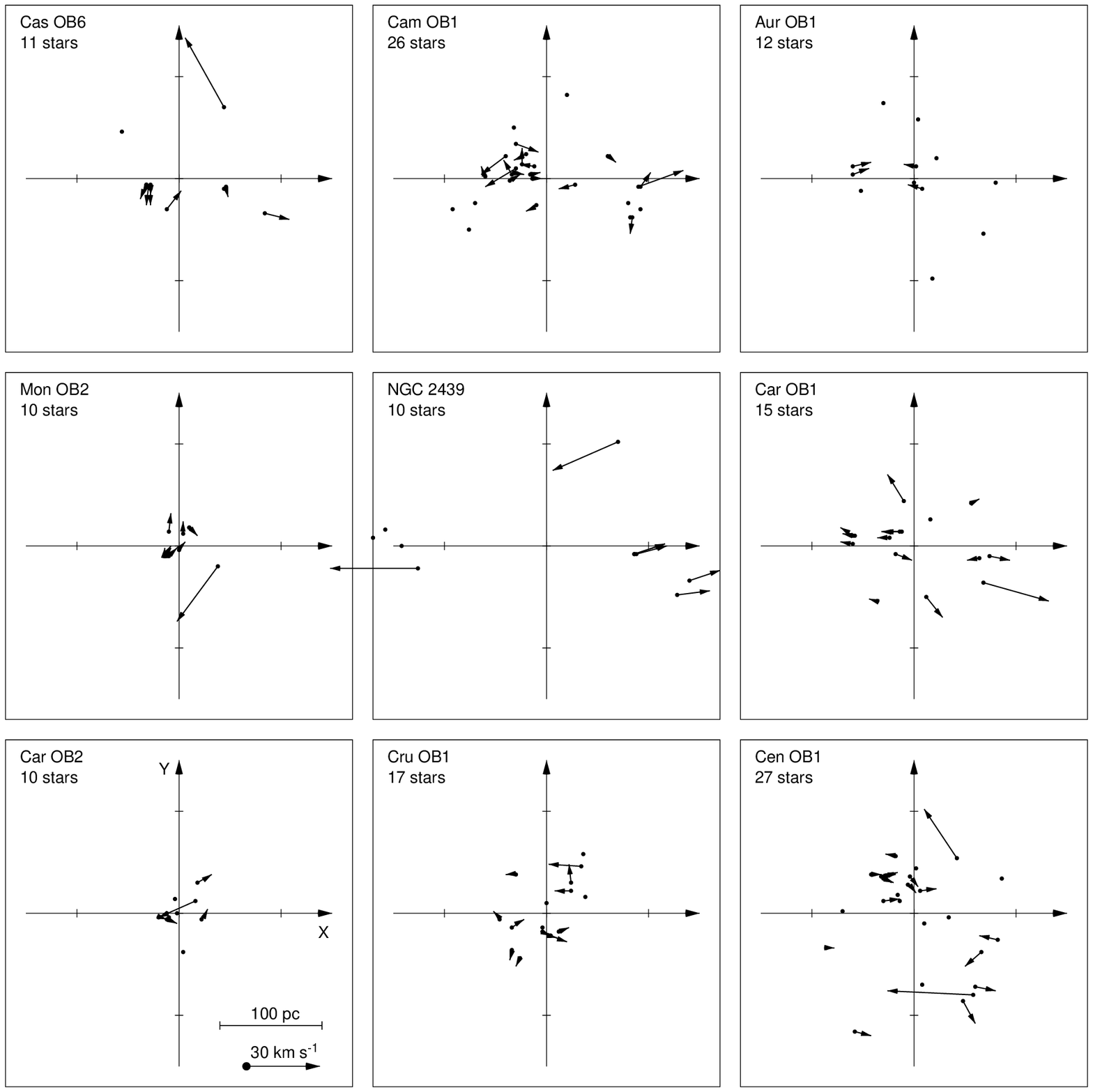}}
\caption{Distribution of the relative velocities $v_l'$ and $v_b'$
(Eq.~\ref{vl'} and ~\ref{vb'})  inside OB-associations with more
than 10 TGAS stars.  See caption to Figure~\ref{expan_all_1} for
more details.} \label{expan_all_2}
\end{figure*}
%------------------------------------------------------------------------------

Figures~\ref{expan_all_1} and~\ref{expan_all_2} show the
distribution of the relative velocities of stars inside
OB-association, $v_l'$ and $v_b'$, calculated with respect to its
average velocities $v_{l0}$ and $v_{b0}$:

\begin{equation}
v_l'= 4.74 r \mu_l  - v_{l0} \label{vl'}
\end{equation}
\begin{equation}
v_b'= 4.74 r \mu_b - v_{b0} \label{vb'}
\end{equation}

Note that we excluded from consideration stars whose relative
velocities exceed  50 km s$^{-1}$, i. e. $|v_l'|>50$ or
$|v_b'|>50$ km s$^{-1}$: BD +57 530A (Per OB1), HD 212043 (Cep
OB2), HD 192445 (Cyg OB3), HD 216878 (Cep OB3), BD +23 3843 (Vul
OB1), HDE 328209 (Ara OB1B), HD 172488 (Sct OB2), and HD 112272 (Cen
OB1). Stars BD +57 530A (Per OB1) and HD 212043 (Cep OB2) have
parallaxes $\pi=12.44\pm0.28$ and $4.91\pm0.63$, which correspond
to distances $r=0.1$ and 0.2 kpc,  so they cannot belong to the
associations Per OB1 and Cep OB2 located at distances $r=1.8$ and
0.7 kpc, respectively. The parallaxes of other aforementioned stars
are consistent with their membership in the corresponding
associations. Generally, they  can be  binary or runaway stars
which can form quite frequently in young clusters \citep[for
example,][]{fujii2011}.

Examination of Figure~\ref{expan_all_1} and
Figure~\ref{expan_all_2} shows that the suspected
expansion/compresssion of some OB-associations is based on the
velocities of a few stars, sometimes only one. For example,
excluding only one of 10--11 stars from the associations  Mon OB2
and NGC 2439 (HD 46573 from Mon OB2 and HD 63423 from NGC 2439)
reduces the absolute values of the parameters of
expansion/compresssion to the noise level. A similar situation is
observed in Cen OB1, which demonstrates compression along
l-coordinate, $p_l=-87\pm33$, at $P>2.5\sigma$ confidence level.
However, excluding four  of 27 stars (HD 115363, HD 114340, HD
110946 and HD 116119) decreases $|p_l|$ to the noise level,
$p_l=-24\pm20$ km s$^{-1}$ kpc$^{-1}$.

We excluded  from our consideration associations  Mon OB2, NGC
2439, and Cen OB1 as having unreliable  parameters of
expansion/compression. The other thee associations (Sgr OB1,  Per
OB1, and Car OB1) with  $\tilde u_l$ and $\tilde u_b$, determined
at $P>2.5$ confidence level demonstrate expansion.

\begin{figure*}
\resizebox{\hsize}{!}{\includegraphics{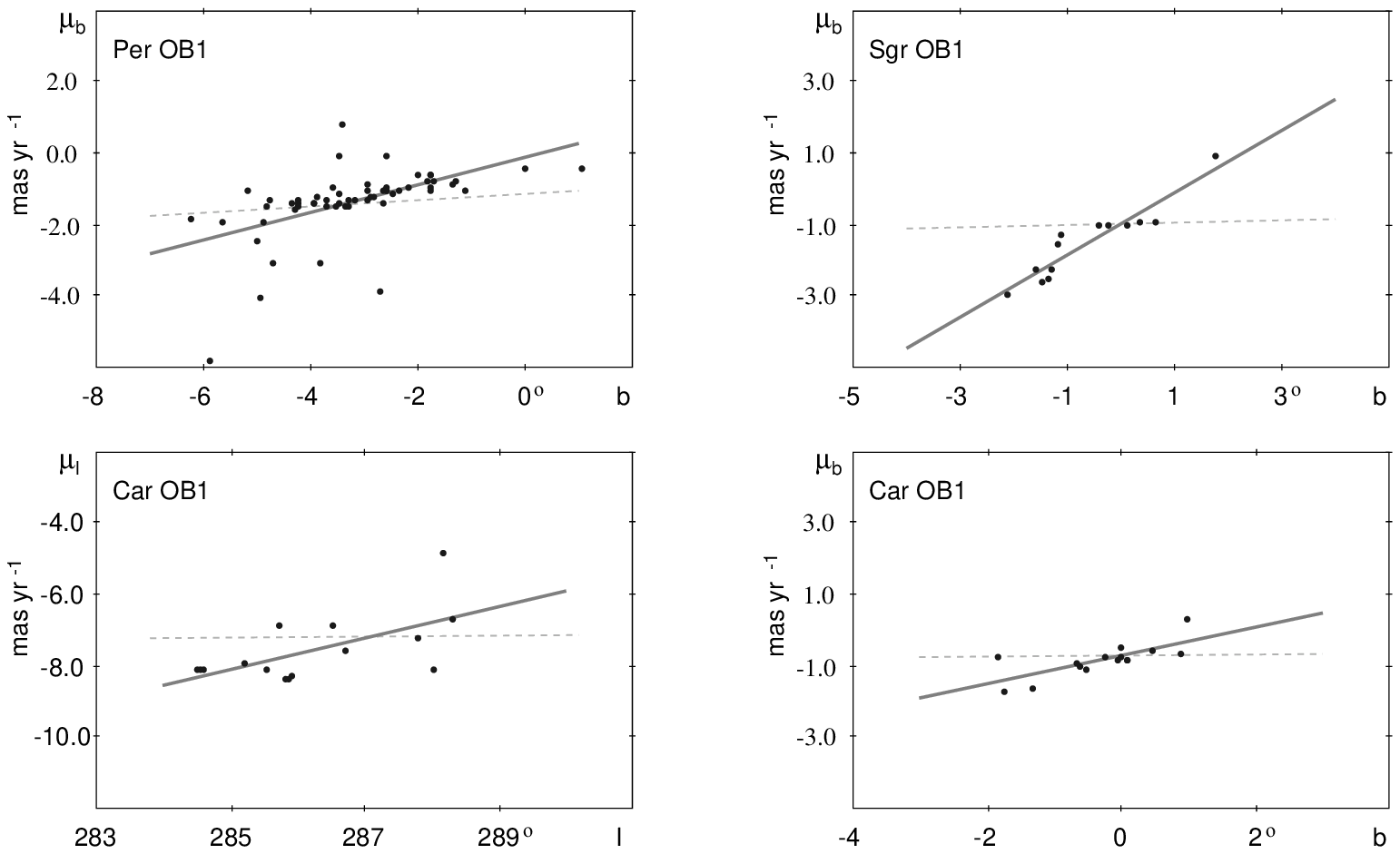}}
\caption{Dependence of the proper motions $\mu_l$ or $\mu_b$ on
the corresponding Galactic coordinates,  $l$   or  $b$,
calculated for TGAS stars in associations Per OB1, Sgr OB1 and Car
OB1.  The solid line indicates the observed linear dependence. The
increase of proper-motion component along the corresponding
coordinate indicates expansion in the chosen direction. The thin
dashed line displays the contribution of spurious expansion due to
line-of-sight velocity $V_r$ (Eq.~\ref{e1}). In all cases
considered  the spurious expansion is conspicuously smaller than
the observed expansion. } \label{ml-l_mb-b}
\end{figure*}

Figure~\ref{ml-l_mb-b} illustrates the expansion  of the three
OB-associations considered. It shows the linear increase  of
proper-motion components, $\mu_l$ or $\mu_b$,   with the
corresponding coordinate, $l$ or $b$, what suggests the expansion
in the chosen direction (a decrease would indicate contraction).
It can be seen that in all cases considered  the spurious
expansion determined by Eq.~\ref{e1} is conspicuously smaller than
the observed expansion.

Generally, the motion of OB-association as a whole in the sky
plane with the  tangential velocity $V_t$  also affects the
velocity distribution inside association by producing spurious
compression in one part of association  and spurious expansion in
another part. The velocity of this spurious expansion/compression
$e_2$ is determined by the following relation:

\begin{equation}
e_2 =  V_t\frac{a^2}{2r^2} \label{e2}
\end{equation}

\noindent with $V_t$ in  the following form:

\begin{equation}
 V_t =  4.74r \cdot \sqrt{\mu_l^2 + \mu_b^2},
\end{equation}

\noindent where $\mu_l$ and $\mu_b$ are given in Table 1.
Formally, the  velocity $e_2$ spuriously increases the velocity
dispersion inside OB-association, but for OB-associations
considered, $e_2$ doesn't exceed $e_2<0.1$ km s$^{-1}$ and its
contribution can be neglected.

%-----------------------    Figure 8   ----------------------------------------
%------------------------------------------------------------------------------

\subsection{Kinematic ages  of OB-associations}

%--------------------------- Table 6 ------------------------------------------
\begin{table}
\caption{Kinematic ages of OB-associations}
 \begin{tabular}{lcc|cc}
 \\[-7pt] \hline\\[-7pt]
Name &$T_l$&$T_b$&$T^*_l$&$T^*_b$\\
& Myr & Myr & Myr & Myr\\
  \\[-7pt] \hline\\[-7pt]
PER OB1   &      --                & $9.5^{+2.5}_{-1.7}$  &            --         & $8.9^{+7.6}_{-3.4}$ \\
[+5pt]
CAR OB1   &  $8.3^{+4.7}_{-2.2}$ &  $8.8^{+4.7}_{-2.2}$  &  $<10$ & $7.4^{+1.1}_{-5.1}$  \\
[+5pt]
SGR OB1   &   --        & $4.1^{+0.5}_{-0.4}$ & -- & $3.3^{+2.7}_{-3.3}$   \\
\hline \end{tabular} \label{kin_age}
\end{table}
%----------------------------------------------------------------------------------

We assume that all stars in an expanding OB-association started
their motion at nearly the same time instant in the past and moved
at nearly constant velocities; hence their distances from the
center $\Delta x$ and $\Delta y$ are proportional to their
observed velocities, $\Delta x=v_l' \cdot t$ and $\Delta y=v_b'
\cdot t$. Though the distances and velocities are different for
different stars, the time interval is supposed to be nearly the
same for all stars in the OB-association considered. We estimate
the so-called kinematic ages of expanding OB-associations by the
formula:

\begin{equation}
 T_l=(p_l\cdot f_v)^{-1},
\label{T_l}
\end{equation}
\begin{equation}
 T_b=(p_b \cdot f_v)^{-1},
\label{T_b}
\end{equation}

\noindent where factor $f_v=3.16/3.09 \cdot 10^{-3}$ transforms
velocities in units of km s$^{-1}$ into kpc Myr$^{-1}$.  The
errors in $T_l$ and $T_b$ are determined by errors in $p_l$ and
$p_b$, respectively.

Table~\ref{kin_age} lists different estimates of the kinematic
expansion ages of the  three OB-associations considered. We can
see that the $T_l$ and $T_b$ ages lie in the interval  4--10 Myr.

Here we ignore the epicyclic motions, which, through Coriolis
forces, change the direction of residual  velocities (those
superimposed on the Galactic rotation). This concerns only
velocities in the Galactic plane. The Galactic rotation curve is
nearly flat, and hence the epicyclic frequency at the solar
distance is $\kappa=\sqrt{2} \Omega_0$ or 43 km s$^{-1}$
kpc$^{-1}$, and one quarter of the epicyclic period amounts to 37
Myr. It is this time interval that takes the Coriolis force to
change the direction of residual velocities by 90$^\circ$. We can
see that the average age of OB-stars, $\sim15$ Myr, amounts to
$\sim40$ percent of one quarter of the epicyclic period. However,
epicyclic motions change the direction of expansion but they do
not prevent it. So the value of $p_l$ is determined by some
combination of the expansion in l-direction and that in the
perpendicular direction -- along the line of sight. Parameter
$p_l$ should rather be viewed as characterizing the expansion in
the Galactic plane. Here we do not consider the expansion along
the line of sight because with TGAS data the accuracy of sky-plane
velocity components is for the first time better than that of
line-of-sight velocities.

%-----------------------    Figure 9   ----------------------------------------
\begin{figure*}
\resizebox{\hsize}{!}{\includegraphics{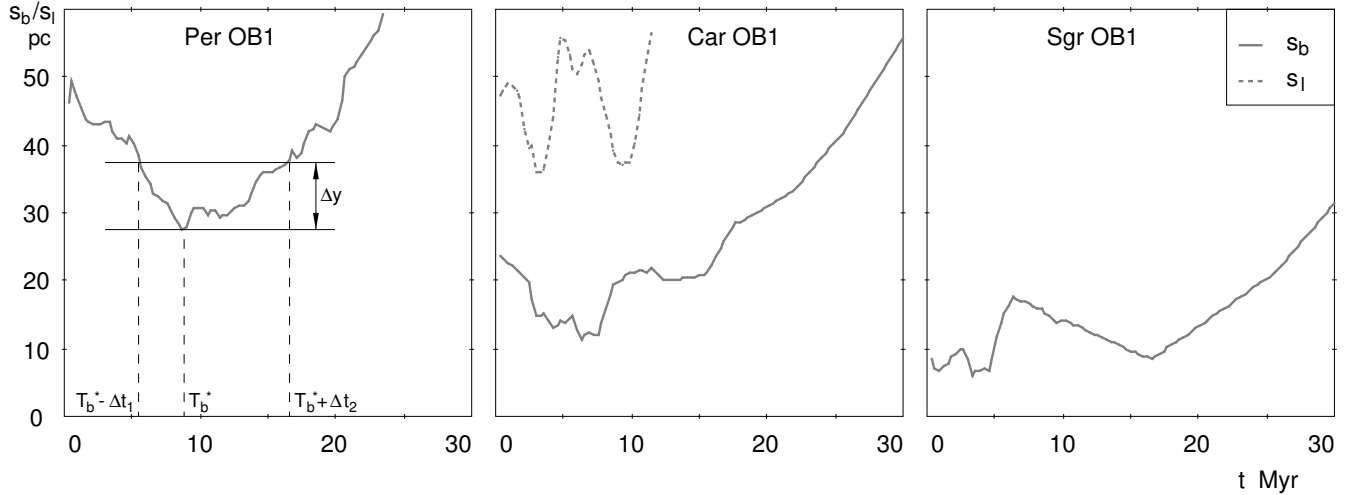}}
\caption{Dependence of the  half-member size of OB-association in
the b- or l-direction,  $s_b$ or $s_l$, on the time $t$ in the
past. Here $s_b$ and $s_l$ are the   lengthes of the minimal
intervals containing 50 percent of members of OB-associations,
$n_\mu$, at the corresponding time moments. The positions of stars
at each time instant were calculated by Eqs.~\ref{past_x}
and~\ref{past_y}. Only OB-associations with well-determined
parameters of expansion are considered.  We illustrate the method
of deriving errors in kinematic ages, $\Delta t_1$ and $\Delta
t_2$, with the example of the Per OB1 association: a horizontal
line $y=s_b(T^*_b)+\Delta y$ intersects the curve $s_b$(t) at two
points with abscissae of $T^*_b-\Delta t_1$ and $T^*_b+\Delta
t_2$, where $\Delta t_1$ and $\Delta t_2$ are error estimates on
the left and right of the minimum.} \label{curves5}
\end{figure*}
%------------------------------------------------------------------------------

There is also another method to determine the time in the past
when OB-associations had minimal sizes \citep[see
also][]{brown1997}. We  trace back the positions of stars in
OB-association using their relative velocities $v_l'$ and $v_b'$:

\begin{equation}
 x(t) = x_0 - v_l' t,
\label{past_x}
\end{equation}
\begin{equation}
 y(t) = y_0 - v_b' t,
\label{past_y}
\end{equation}

\noindent  where $x_0$ and $y_0$ are the observed coordinates of
TGAS stars with respect to the centre of OB-association, and
compute the sizes of the most compact parts of OB-association at
different time instants.   The  sizes, $s_l$(t) and $s_b$(t),
containing 50 percent of TGAS stars of the corresponding
OB-association are calculated in the following way.  We sort the
coordinates of association member stars, $x$ and $y$, in the
ascending order and seek the subset $S = \{j+1,j+2.... j+k \}$
containing $k = n_\mu/2$ stars (where $n_\mu$ is the number of
TGAS stars in OB-association) having the smallest span in the
corresponding coordinate, $x_{j+k}$-$x_{j+1}$ and
$y_{j+k}$-$y_{j+1}$. These minimal spans in the X- and Y-direction
give the values of $s_l$(t) and $s_b$(t), respectively. That is
done to mitigate the effect of individual stars.

Figure \ref{curves5} shows the dependence of $s_b$ and $s_l$ on
the time  $t$ in the past.  We consider only  OB-associations Per
OB1, Car OB1 and Sgr OB1 with well-determined parameters $\tilde
u_l$ or $\tilde u_b$. The curves $s_b$(t) and $s_l$(t) are often
indented and some of them have several local minima, which are due
to abrupt changes of the estimated size due to inclusion/exclusion
of individual stars  in the half-member minimal-span subset. We
determine the kinematic expansion ages of OB-associations, $T^*_b$
and $T^*_l$, as the time instants $t$ in the past corresponding to
the minimal values of $s_b$ and $s_l$, respectively (Fig.~
\ref{curves5}). However, in some cases we can give only an upper
estimate for the kinematic age.

Table~\ref{kin_age} lists the inferred $T^*_l$ and $T^*_b$ ages
and their errors,  which are calculated through the distances,
$\Delta x$ and $\Delta y$, in the l- and b-direction which stars
pass due to their errors in proper motions, $\varepsilon_{\mu l}$
and $\varepsilon_{\mu b}$, during the time intervals $T^*_l$ and
$T^*_b$, respectively:

\begin{equation}
\Delta x = 2\cdot 4.74 r  \varepsilon_{\mu l} \cdot T^*_l f_v,
\label{err_x}
\end{equation}
\begin{equation}
\Delta y = 2\cdot 4.74 r  \varepsilon_{\mu b} \cdot T^*_b f_v,
\label{err_y}
\end{equation}

\noindent where $\varepsilon_{\mu l}$ and $\varepsilon_{\mu b}$
are in mas yr$^{-1}$ but  $\Delta x$, $\Delta y$ and $r$ are in
kpc. Factor two reflects possible changes of the sizes of
OB-association at  left and right sides of the distribution. The
average errors in proper motions, $\varepsilon_{\mu l}$ and
$\varepsilon_{\mu b}$, lie in the range 0.05--0.07 mas yr$^{-1}$.
The distances $\Delta x$ and $\Delta y$ increase with increasing
the distance to the OB-association, $r$, and with increasing
kinematic ages, $T^*_l$ or $T^*_b$.  They have the following
values: $\Delta y=10$ pc (Per OB1),  $\Delta x\approx\Delta
y=7\textrm{--}10$ pc (Car OB1) and  $\Delta y=4$ pc (Sgr OB1).

To transform errors in distances, $\Delta x$ and $\Delta y$, into
errors in time intervals, $\Delta t_1$ and $\Delta t_2$, we use
following expressions:

\begin{equation}
 s_l(T^*_l)+\Delta x=s_l(T^*_l-\Delta t_1)=s_l(T^*_l+\Delta t_2),
\label{err_tl}
\end{equation}
\begin{equation}
 s_b(T^*_b)+\Delta y=s_b(T^*_b-\Delta t_1)=s_b(T^*_b+\Delta t_2),
\label{err_tb}
\end{equation}

\noindent where $\Delta t_1$ and $\Delta t_2$ are the errors
estimated on the left and right of the minimum on the curves
$s_l$(t) or $s_b$(t). To illustrate this method we chose
association Per OB1 and drew a horizontal line with the ordinate
$y=s_b(T^*_b)+\Delta y$ which intersects the curve $s_b$(t)  at
two points with abscissae of $T^*_b-\Delta t_1$ and $T^*_b+\Delta
t_2$ (Fig.~\ref{curves5}).

Table~\ref{kin_age} shows that  the kinematic ages of
OB-associations obtained by two different methods, $T_l$ and
$T^*_l$ or $T_b$ and $T^*_b$ agree with each other  within the
errors. The kinematic ages, $T_l$, $T^*_l$, $T_b$ and $T^*_b$,
determined by both methods for three OB-associations considered
lie in the range 3--10 Myr. These estimates  do not exceed the
main-sequence ages of O--B2 stars, $T_s<30$ Myr
\citep{bressan2012}.

\subsection{Minimal sizes of  OB-associations in the past}

The second method  is also interesting because it gives  us  the
estimates of the minimal sizes of OB-associations corresponding to
the time instants $T^*_l$ and $T^*_b$ (Fig.~\ref{curves5}). The
minimal sizes of the most compact parts of OB-associations
containing 50 percent of star-members, $s_l$($T^*_l$) and
$s_b$($T^*_b$), appear to be 27 pc for Per OB1,  11--36 for Car
OB1,  and 6 pc for Sgr OB1.

Figure~\ref{curves5} shows that sizes of Per OB1  and Car OB1 in
latitude directions, $s_b$, were nearly twice smaller in the past
than nowadays. The formula for the virial mass (Eq.~\ref{m_vir})
includes the radius of the system $a$ referred to the epoch of
star formation and we therefore should correct their virial masses
$M_{vir}$. Assuming that the contributions to $a^2$ from the
coordinates $x$ and $y$ are nearly the same, we must reduce
$M_{vir}$ of Per OB1 and Car OB1 by the factor
$(1+0.25)/2)^{1/2}=0.79$ and  upscale star-formation efficiency
$\epsilon$ by a factor of 1.26 to the new values of 4.4 and 1.8
percent, respectively.

We can see that the minimal sizes of  the  Per OB1 and Car OB1
associations lie in the range  11--27 pc and are consistent with
diameters of giant molecular clouds, 10--80 pc. As for Sgr OB1,
the situation is unclear here: we see too quick expansion in very
small area. Moreover, it contains only 13 stars with known TGAS
proper motions and errors of their $\mu_b$ are the highest,
$\varepsilon_{\mu b}=0.07$ mas yr$^{-1}$, among stars in the
OB-associations considered.

Possibly, the quick expansion of Sgr OB1, as well as the large
velocity dispersion in Cyg OB7 and Cyg OB8 have the same origin --
larger errors of proper motions in fields rich in bright stars.

%-----------------------    Figure 10   ----------------------------------------
\begin{figure}
\resizebox{\hsize}{!}{\includegraphics{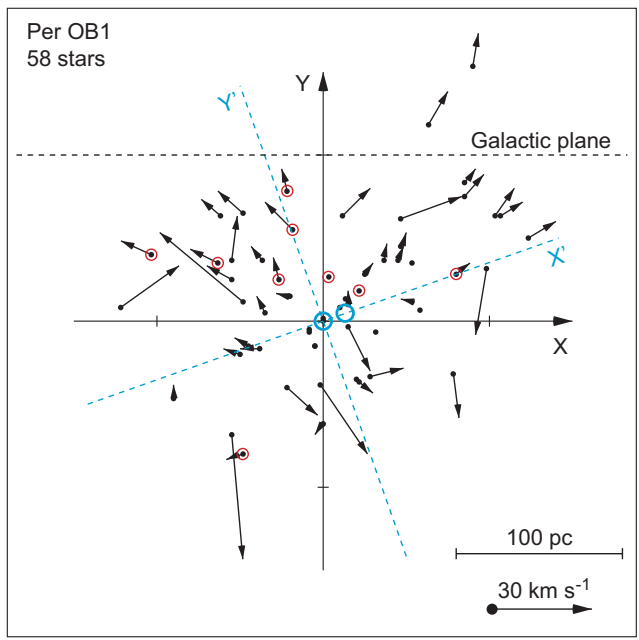}} \caption{
Distribution of relative velocities $v_l'$ and $v_b'$ in the Per
OB1 association. Axes $X$ and $Y$ are directed toward increasing
values of Galactic coordinates $l$ and $b$, respectively. Two open
clusters NGC 869 ($l=134.63^\circ$, $b=-3.74^\circ$) and NGC 884
($l=135.07^\circ$, $b=-3.59^\circ$) lying practically in the
centre of the OB-association are shown as circles (colored blue in
electronic edition). The line connecting them (the dashed line) is
tilted by $20^\circ$ to the Galactic plane. The  expansion of Per
OB1 ($p= 136\pm25$ km s$^{-1}$ kpc$^{-1}$) is the fastest in the
direction nearly perpendicular to this line.  The supergiants of
type K and M are outlined by circles (colored red in electronic
edition). The position of the Galactic plane is shown by the black
dashed line.} \label{per_ob1}
\end{figure}
%------------------------------------------------------------------------------

\subsection{Expansion of Per OB1}

Let us consider the expansion of  the Per OB1 association  in more
detail (Fig. ~\ref{per_ob1}). It  exhibits significant expansion
along  b-coordinates, $p_b=103\pm22$ km s$^{-1}$ kpc$^{-1}$
(Table~\ref{par_expansion}). However, the expansion can be even
faster if we tilt the coordinate axes ($X$,~$Y$) by the angle
$\alpha=26^\circ$ with respect to the Galactic plane. In the new
reference frame ($X'$,~$Y'$) the expansion parameter has the
values of $p_y'=136\pm25$ kpc$^{-1}$ corresponding to the velocity
7.5  km s$^{-1}$. Interestingly, the new axis $X'$ is nearly
parallel to the line connecting two rich open clusters NGC 869
($l=134.63^\circ$, $b=-3.74^\circ$) and NGC 884 ($l=135.07^\circ$,
$b=-3.59^\circ$) also known as the double cluster $h$ and $\chi$
Persei lying practically at the centre of OB-association Per OB1.
So the maximal expansion happens in the direction nearly
perpendicular to the main plain of the double cluster.

We  subdivide stars of Per OB1 into two groups:  red supergiants
of spectral type K and M (nine objects with TGAS proper motions)
and stars of type O-B (49 objects). The ages of red supergiants
must be less than 50 Myr while those of the main-sequence O--B2
stars must be  less than 30 Myr \citep{bressan2012}. However, we
do not see difference in their kinematical behavior: both groups
take part in the expansion along l- and b-directions. The
parameters of expansion obtained  for red supergiants exclusively
are $p_l=75\pm15$ and $p_b=61\pm14$ km s$^{-1}$ kpc$^{-1}$, while
those derived for OB-stars only are $p_l=35\pm18$ and
$p_b=109\pm26$ km s$^{-1}$ kpc$^{-1}$. These parameters are
consistent within $1.2\sigma$. Probably, red supergiants and OB
stars observed in Per OB1 have very close ages and take part in
the same expansion.

\citet{cappa2000} notice  that there is a lot of neutral hydrogen
and  no molecular gas in the vicinity of  the Per OB1 association.
They found a lot of HI bubbles surrounding massive WR and Of stars
there.

Note that the evolutionary ages of clusters $h$ and $\chi$ Persei
are comparable to those of stars in OB-associations, 10 Myr
\citep{dias2002, kharchenko2013} and the infered expansion age of
the association. Possibly, the expansion of Per OB1 and formation
of double stellar cluster $h$ and $\chi$ Persei in its centre have
the same cause -- the quick loss of gas in their parent giant
molecular cloud.

As for  double clusters, \citet{kroupa2001}    find multiple
clusters in some of their  simulations of the formation of bound
clusters inside the expanding OB-associations. Though their models
are very compact, $\sim 10$ pc, the physical mechanism considered
can be universal. The authors explain the formation of double
clusters in their models by gravitational interactions between
neighbour stars, which cause the redistribution of the velocity
dispersion from the radial direction into the azimuthal one.  This
mechanism can lead to the formation of a subsystem with non-zero
angular momentum about the origin of the expanding flow.

\section{Conclusions}

We have studied the kinematics of OB-associations identified by
\citet{blahahumphreys1989} using the stellar proper motions from
the  TGAS catalog \citep{michalik2015}. The average difference  in
the median proper potions of OB-associations obtained with TGAS
and Hipparcos data is 0.67 mas yr$^{-1}$, which is
comparable to  the average  error of Hipparcos proper motions,
0.916 mas yr$^{-1}$, for stars of OB-associations.

Generally, the results based on TGAS and Hipparcos catalogs
demonstrate a good agreement. The parameters of the Galactic
rotation curve obtained with TGAS (Table~\ref{rot_curve}) and
Hipparcos \citep{melnikdambis2009} proper motions agree within the
errors. The same can be said about residual velocities
($V_{res}=V_{obs} - V_{rot}-V_{ap}$ ) of OB-associations in
stellar-gas complexes identified by \citet{efremov1988}. The rms
deviation of the velocities of OB-associations from the Galactic
rotation curve amounts $\sigma_0=7.2\textrm{--}7.5$ km s$^{-1}$.
This value is nearly the same whether computed with TGAS or
Hipparcos data, suggesting  that the residual velocities of 7--8
km s$^{-1}$ are real and not an artifact.

When computed with TGAS  data, the rms deviation of the velocities
of OB- associations in the z-direction is $\sigma_{vz}=2.7$ km
s$^{-1}$, which is considerably less than  the scatter obtained
with Hipparcos data, $\sigma_{vz}=5.0$ km s$^{-1}$.

We  found a considerable decrease, by a factor of 0.4, in the
velocity dispersions inside 18 OB-associations containing more
than 10 TGAS stars compared to the values derived with Hipparcos
data. The average one-dimensional velocity dispersion inside
OB-associations computed with TGAS catalog data is
$\overline{\sigma_v}=3.9$ km s$^{-1}$.

The effective contribution from binary OB-stars into the velocity
dispersion $\sigma_v$  inside OB-associations is $\sigma_b=1.2$ km
s$^{-1}$.

The virial $M_{vir}$ and stellar $M_{st}$ masses of 18
OB-associations considered have the median values of 7.1~10$^5$
and 9.0~10$^3$ M$_\odot$, respectively. Generally, the virial mass
exceeds the stellar mass by more than 70 times. This fact together
with assumption that the volumes of OB-associations do not contain
a lot of dense  gas supports the conclusion that they are unbound
objects.

The star-formation efficiency $\epsilon$ in OB-associations varies
from 0.1 to 13.6 percent with the median of $\epsilon=2.1$
percent, which agrees with  other estimates \citep{myers1986,
evans2009, garcia2014}.

We determined the tidal radii of OB-associations under two
assumptions: (1) OB-associations include neutral gas and its mass
is proportional to the volume of the association, so that the
present-day mass of the OB-association consists of masses of the
gas and stellar components; (2) the masses of OB-associations are
determined by their stellar component exclusively. In the first
case, the median tidal radius appears to be 40 pc, suggesting that
27 percent of member stars of OB-associations are located outside
their tidal radii. In the second case, the tidal radii range from
14 to 36 pc with the median value of of 26 pc, implying that  39
percent of stars of OB-associations must lie outside their tidal
radii. Thus, $\sim 1/3$  of stars of OB-associations must lie
outside their tidal radius.

The high accuracy of TGAS proper motions allowed us to find the
expansion inside  the OB-associations Per OB1 and  Car OB1
supporting the idea that they are  unbound objects. The velocity
of expansion amounts, on average, to 6.3 km s$^{-1}$
(Table~\ref{par_expansion}).

We corrected the observed specific velocities of
expansion/compression, $u_l$ and $u_b$, for the effect of spurious
expansion/compression caused by  the motion of OB-association as a
whole with the line-of-sight velocity $V_r$ (Eq.~\ref{e1}). Only
cases with well-determined corrected velocities, $\tilde u_l$ and
$\tilde u_b$, are considered (Table~\ref{par_expansion}).

The kinematic ages of the expanding OB-associations Per OB1 and
Car OB1 (time moments in the past when they had their minimal
sizes) calculated by two methods are  7--10 Myr. The minimal
diameter of the most compact parts of these OB-associations in the
past lies in the range 11--27 pc which is comparable to the
diameters of Galactic giant molecular clouds, 10--80 pc
\citep{sanders1985}.

We studied in more detail the expansion of the Per OB1
association. The direction of its fastest expansion ($Y'$ in
Fig.~\ref{per_ob1}) appears to be  tilted by $26^\circ$ with
respect to the b-direction and is nearly perpendicular to the
plane determined by the positions of the two rich open clusters,
$h$ and $\chi$ Persei,  lying in the centre  of the association.
We found no significant difference between the parameters of
expansion derived for OB-stars and red supergiants in this
association.

\section{acknowledgements}

We thank  A.G.A. Brown for fruitful discussion. We  thank A.S.
Rastorguev and O.K. Sil'chenko for useful remarks and suggestions.
This work has made use of data from the European Space Agency
(ESA) mission {\it Gaia} (\verb"https://www.cosmos.esa.int/gaia"),
processed by the {\it Gaia} Data Processing and Analysis
Consortium (DPAC,
\verb"https://www.cosmos.esa.int/web/gaia/dpac/consortium").
Funding for the DPAC has been provided by national institutions,
in particular the institutions participating in the {\it Gaia}
Multilateral Agreement. This research has made use of the VizieR
catalogue access tool, CDS, Strasbourg, France. The original
description of the VizieR service was published  by
\citet{ochsenbein2000}. Analysis of  observational data was
supported by Russian Scientific Foundation grant no. 14-22-00041.

\end{document}